\documentclass[12pt]{article} 

\usepackage{color}
\usepackage{a4wide}
\usepackage{amssymb}
\usepackage{graphicx}

\def\gtap{\raisebox{-.4ex}{\rlap{$\sim$}} \raisebox{.4ex}{$>$}} 
\def\beq{\begin{equation}} 
\def\eeq{\end{equation}} 
\def\barr{\begin{array}}
\def\earr{\end{array}}
\def\dis{\displaystyle}
\def\gev{\, {\rm GeV}} 
\def\tev{\, {\rm TeV}} 

\begin{document} 
 
\vskip 30pt 
 
\begin{center} 
{\Large \bf Deciphering Universal Extra Dimension from the top quark 
signals at the CERN LHC} \\
\vspace*{1cm} 
{ {\sf Debajyoti Choudhury${}^1$}, {\sf Anindya Datta${}^{2}$} and
{\sf Kirtiman Ghosh${}^{2}$} 
} \\ 
\vspace{10pt} 
{\small ${}^{1)}$ {\it Department of Physics and Astrophysics, 
University of Delhi, Delhi 110007, India.} \\[1ex]
   ${}^{2)}$ {\it Department of Physics, University of Calcutta, \\
92 Acharya Prafulla Chandra Road, Kolkata 700009, India.}}  \\
\normalsize 
\end{center} 
 
\begin{abstract} 
\noindent
Models based on Universal Extra Dimensions predict Kaluza-Klein (KK)
excitations of all Standard Model (SM) particles. We examine the pair
production of KK excitations of top- and bottom-quarks at the Large Hadron
Collider. Once produced, the KK top/bottom quarks can decay to $b$-quarks,
leptons and the lightest KK-particle, $\gamma_1$, resulting in 2
$b$-jets, two opposite sign leptons and missing transverse momentum, 
thereby mimicing top-pair production.  We show that, with a proper choice of
kinematic cuts, an integrated luminosity of  100 fb$^{-1}$ would 
allow a discovery for an inverse radius upto $R^{-1} = 750$ GeV.

\vskip 5pt \noindent 
\texttt{Key Words:~~Universal Extra Dimension, LHC,  etc.} 
\end{abstract}

\renewcommand{\thesection}{\Roman{section}} 
\setcounter{footnote}{0} 

\section{Introduction}

The twin primary goals of Large Hadron Collider (LHC), expected to
start operating by the end of 2009 are to understand the
mechanism for electro-weak symmetry breaking (EWSB) as well as 
uncover any 
new dynamics that may be operative at the scale of a few TeVs.
Two classes of theories command special
attention at  future experiments like the LHC: supersymmetric theories 
on the one hand and those invoking one or more 
{\it extra space-like dimensions} on the other. 
It should be noted that these two ideas (supersymmetry and 
additional dimensions) are by no means mutually exclusive.

Extra-dimensional theories can be subdivided into two main
classes. The first corresponds to those wherein the Standard Model
(SM) fields are confined to a (3 + 1) dimensional subspace of the full
manifold. We shall not concern ourselves with such models.
  The second class comprises models
wherein some or all of the SM fields can access the extended space-time
manifold \cite{acd}, whether fully or partially.
  Such TeV scale extra-dimensional scenarios could
lead to a new mechanism of supersymmetry breaking \cite{antoniadis1},
relax the upper limit of the lightest supersymmetric neutral Higgs
\cite{relax}, address the issue of fermion mass hierarchy from a
different perspective \cite{Arkani-Hamed:1999dc}, provide a cosmologically
viable dark matter candidate \cite{Servant:2002aq}, interprete the
Higgs as a quark composite leading to a successful EWSB
without the necessity of a fundamental scalar or Yukawa
interactions \cite{Arkani-Hamed:2000hv}, and lower the unification
scale down to a few TeVs \cite{dienes,dienes2,blitz}. Our concern here
is a specific and particularly interesting 
framework, called the Universal Extra Dimension (UED)
scenario, characterized by a single flat extra dimension, compactified
on an $S^1 /Z_2$ orbifold, which is accessed by all the SM particles
\cite{acd}. From a 4-dimensional viewpoint, every field will then have
an infinite tower of Kaluza-Klein (KK) modes, the zero modes being
identified as the corresponding SM states.

The key feature of the UED Lagrangian is that the momentum in the
universal fifth direction is conserved. From a 4-dimensional
perspective, this implies KK number conservation. However, boundary
conditions break this symmetry, leaving behind only a conserved KK
parity, defined as $(-1)^n$, where $n$ is the KK number. This 
discrete symmetry ensures 
that the lightest KK particle is stable (hence being a natural candidate
for the Dark Matter particle~\cite{Servant:2002aq}) 
and that the level-one KK-modes would be produced only in
pairs.  This also ensures that the KK modes do not affect electroweak
processes at the tree level. And while they do contribute to higher
order electroweak processes, in a loop 
they appear only in pairs resulting in a
substantial suppression of such contributions, thereby allowing for
relatively smaller KK-spacings.  In spite of the infinite multiplicity
of the KK states, the KK parity ensures that all electroweak
observables are finite (up to one-loop)\cite{db}\footnote{The observables start
  showing cutoff sensitivity of various degrees as one goes beyond
  one-loop or considers more than one extra dimension.}, and a
comparison of the observable predictions with experimental data yields
bounds on the compactification radius $R$.  Constraints on the UED
scenario from the measurement of the anomalous magnetic moment
of the muon \cite{nath}, flavour changing neutral
currents \cite{chk,buras,desh}, $Z \to b\bar{b}$ decay \cite{santa},
the $\rho$ parameter \cite{acd,appel-yee}, several other electroweak
precision tests \cite{ewued}, all conclude that $R^{-1}~\gtap~300$ GeV.

The fact that such a small value for $R^{-1}$ (equivalently, small KK
spacings) is still allowed, renders collider search prospects very
interesting both in the context of hadronic \cite{collued} and leptonic
\cite{ILC} colliders.

In this work, we study a very specific signal of the
universal extra dimension at the Large Hadron Collider, culminating in
a final state that would be very well-studied at the LHC irrespective
of the existence of any such models.
 To be precise, we consider the pair production of the first KK-top
 and KK-bottom quarks at the LHC. While decaying, these may go into a
 pair of $b$-jets accompanied by a pair of opposite-signed leptons as
 well as missing transverse momentum ($p_T\hspace*{-1em}/\hspace*{1ex}$). 
This specific
 final state also arises from the pair production and decay of top
 quarks and indeed forms a bedrock of top-quark study. Thus, the same
 data sample can be utilised for exploring the signals of UED.

The rest of paper is organised as follows. 
In the next section, we briefly discuss the universal
extra-dimensional scenario with emphasis on its mass spectrum.  In the
subsequent section, we will discuss the production and decay of the
first KK-top and KK-bottom quarks. 
We also compare the decay signal with that of SM
top quarks, which will be produced copiously at the LHC. We will
discuss the kinematic cuts which can seperate the UED signal from that
of SM top quarks. Finally, we  summarise in section IV.

\section{Universal Extra Dimension}\label{ued} The extra dimension is
compactified on a circle of radius $R$ with a $Z_2$ orbifolding
defined by identifying $y \to -y$, where $y$ denotes the fifth
(compactified) coordinate. The orbifolding is crucial in generating
{\it chiral} zero modes for fermions. Each component of a
5-dimensional field must be either even or odd under the orbifold
projection. After integrating out the compactified dimension, the
effective 4-dimensional Lagrangian can be written in terms
of the respective zero modes and the KK excitations.  It is instructive 
to take a glance at the KK mode expansions of these fields.  Defining
\beq
{\cal C}_n \equiv \sqrt{\frac{2}{\pi \, R}} \, \cos \, \frac{n \, y}{R} 
\qquad 
{\rm and}
\qquad
{\cal S}_n \equiv \sqrt{\frac{2}{\pi \, R}} \, \sin \, \frac{n \, y}{R} \ ,
\eeq
the KK expansions are given by
\beq
\label{fourier}
\barr{rcl}
A_{\mu}(x,y)&=&\dis \frac{1}{\sqrt{2}} \, A_{\mu}^{(0)}(x) \, {\cal C}_0 
               + \sum^{\infty}_{n=1} A_{\mu}^{(n)}(x) \, {\cal C}_n \ ,
\\[1ex]
A_5(x,y) & = & \dis \sum^{\infty}_{n=1}A_5^{(n)}(x)\, {\cal S}_n  \ ,
\\[1ex]
\phi(x,y)&=& \dis \frac{1}{\sqrt{2}} \, \phi^{(0)}(x) {\cal C}_0 
    + \sum^{\infty}_{n=1}\phi^{(n)}(x)  \, {\cal C}_n \ ,
\\[2.5ex]
\mathcal{Q}_{i}(x,y)&=&\dis \frac{1}{\sqrt{2}} \, \mathcal{Q}_{i}(x) \,
                             {\cal C}_0  
    + \sum^{\infty}_{n=1}\left[ \mathcal{Q}^{(n)}_{iL}(x) \, {\cal C}_n +
                        \mathcal{Q}^{(n)}_{iR}(x) \, {\cal S}_n \right] \ ,
 \\[2.5ex]
\mathcal{U}_{i}(x,y)&=&\dis \frac{1}{\sqrt{2}} \, \mathcal{U}_{i}(x) \,
                             {\cal C}_0  
    + \sum^{\infty}_{n=1}\left[ \mathcal{U}^{(n)}_{iR}(x)\, {\cal C}_n
                       + \mathcal{U}^{(n)}_{iL}(x)\, {\cal S}_n \right] \ ,
 \\[2.5ex]
\mathcal{D}_{i}(x,y)&=&\dis \frac{1}{\sqrt{2}} \, \mathcal{D}_{i}(x) \,
                             {\cal C}_0  
    + \sum^{\infty}_{n=1}\left[ \mathcal{D}^{(n)}_{iR}(x)\, {\cal C}_n
                        + \mathcal{D}^{(n)}_{iL}(x)\, {\cal S}_n \right] \ ,
\earr
    \label{mode_expan}
\eeq
where $i=1\dots3$ denotes generations and the fields $\mathcal{Q}_i$,
$\mathcal{U}_i$, and $\mathcal{D}_i$ describe the 5-dimensional quark
weak-doublet and singlet states respectively.  The zero modes thereof
are identified with the 4-dimensional chiral SM quark states.  The
complex scalar field $\phi (x,y)$ and the gauge boson $A_\mu(x,y)$ are
$Z_2$ even fields with their zero modes identified with the SM scalar
doublet and SM gauge bosons respectively. On the contrary, the field
$A_5(x,y)$, which is a real scalar transforming in the adjoint
representation of the gauge group, does not have any zero mode.  The
KK expansions of the lepton fields are analogous to those for the quarks
and are not shown for the sake of brevity.

\subsection{Radiative corrections to the KK-masses}\label{rKK}

As is well known, the tree-level mass of a level-$n$ KK component is
given by \beq m_n^2 = m_0^2 + \frac{n^2}{R^2} \ , \eeq 
$m_0$
is the mass associated with the corresponding SM field.  Now, as a
zeroth approximation, one may treat all the SM particles, apart from
the top quark, the weak gauge bosons and the Higgs boson, to be nearly
massless.  This implies, that, at each KK-level, the different states
are approximately degenerate. An immediate consequence would be the
quasi-stability of many of the KK-fields, leading to possibly
spectacular signatures in a collider environment. This, however, is
only the leading order approximation and quantum corrections lift this
degeneracy. Apart from the usual radiative corrections that we expect
in a Minkowski-space field theory, there are additional corrections
accruing from the fact of the fifth direction being a compact one.  We
briefly review these next.

\begin{itemize}
\item Bulk corrections:\\
These arise
due to the winding of the internal loop (lines) around the compactified
direction\cite{radi_matchev}, 
and are nonzero (and finite) only for the gauge boson KK-excitations.
For the first level KK-modes, the bulk corrections are given by  
\beq
\barr{rcl}
\delta\, (m_{B^n}^2) &=& \dis -\frac{39 \,\zeta(3) \, a_1}{2 \, \pi^2 \, R^2} \ ,
 \\[2ex]
\delta\, (m_{W^n}^2) &=&  \dis  -\frac{5\,\zeta(3) \, a_2}{2 \, \pi^2 \, R^2}\ ,
\\[2ex]
\delta\, (m_{g^n}^2) &=& \dis  -\frac{3\,\zeta(3) \, a_3}{2 \, \pi^2 \, R^2}\ ,
\earr
\label{delta0}
\eeq
where $a_i \equiv g_i^2 / 16 \, \pi^2 \ , i = 1\dots 3$ with $g_i$ 
denoting the respective gauge coupling constants. 
The vanishing of these corrections in the limit $R \rightarrow \infty$
reflects 
the removal of the compactness of the fifth direction and hence the 
restoration of full five-dimensional Lorentz invariance.

\item Orbifold corrections:\\ 
The very process of orbifolding introduces a set of special (fixed)
points in the fifth direction (two in the case of $S^1/Z_2$
compactification). This clearly violates the five-dimensional Lorentz
invariance of the tree level Lagrangian.  Unlike the bulk corrections,
the boundary corrections are not finite, but are logarithmically
divergent\cite{radi_matchev}.  They are just the counterterms of the
total orbifold correction, with the finite parts being completely
unknown, dependent as they are on the details of the ultraviolet
completion.  Assuming that the boundary kinetic terms vanish at the cutoff
scale $\Lambda$ ($=20$ TeV here) the corrections from the boundary terms,
at a renormalization scale $\mu$ would obviously be proportional to 
$L_0 \equiv \ln (\Lambda^2 /\mu^2) $. Denoting $m_n({\cal A})$ to be the 
tree-level mass of the $n$-th KK-component of a SM field ${\cal A}$, we 
have\cite{radi_matchev}

\beq
\barr{rcl}
\bar{\delta}\, m_{Q^n} &=& \dis 
m_n \,\left(3\, a_3
+ \frac{27}{16}\,a_2 + \frac{a_1}{16}
\right) \,L_0 \ ,
\\[2ex]
\bar{\delta}\, m_{u^n} &=& m_n \,\left(3\, a_3
  + a_1
\right) \, L_0 \ ,
\\[2ex]
\bar{\delta}\, m_{d^n} &=& \dis m_n \,\left(3\, a_3
  + \frac{a_1}{4}
\right) \, L_0 \ ,  \\[2ex]
\bar{\delta}\, m_{L^n} &=& \dis m_n \,\left(
 \frac{27}{16}\,a_2 + \frac{9}{16}\,a_1
\right) \, L_0 \ ,  \\[2ex]
\bar{\delta}\, m_{e^n} &=& \dis
 \frac{9\,a_1}{4}
\,  m_n \,L_0 \ , \\[2ex]
\bar{\delta}\, (m_{B^n}^2) &=& \dis \frac{-a_1}{6}\, m_n^2\,  L_0 
 \\[2ex]
\bar{\delta}\, (m_{W^n}^2) &=& \dis \frac{15\, a_2}{2}
\, m_n^2\, L_0 \ , \\[2ex]
\bar{\delta}\, (m_{g^n}^2) &=& \dis \frac{23\, a_3}{2} \,  m_n^2 \,
L_0 \ , \\[2ex]
\bar{\delta}\, (m_{H^n}^2) &=& \dis m_n^2 \, \left(\frac{3}{2}\, a_2   
+ \frac{3}{4}\, a_1 - \frac{\lambda_H}{16\pi^2} \right)
L_0 + \overline{m}_H^2\ , 
\earr
\label{delta1}
\eeq
where the boundary term for the Higgs scalar, namely $\overline{m}_H^2$,
 is taken to be vanishing. 
\end{itemize}

The KK-excitations of the neutral electroweak gauge bosons mix in a
fashion analogous to their SM counterparts and the mass eigenstates
and eigenvalues of the KK `photons' and `$Z$' bosons are obtained by
diagonalizing the corresponding mass squared matrices.  In the $(B_n,
W^3_n)$ basis, the latter reads 
\[
\left( \barr{cc} 
\dis
\frac{n^2}{R^2}+ \hat{\delta} m_{B^n}^2 + \frac{1}{4}g_1^2 v^2 
& \dis
\frac{1}{4}g_1 g_2 v^2 
\\[2ex] 
\dis 
\frac{1}{4}g_1 g_2 v^2 & \dis
\frac{n^2}{R^2}+ \hat{\delta} m_{W^n}^2 +\frac{1}{4}g_2^2 v^2 \earr
\right), 
\]
where $\hat{\delta}$ represents the total
one-loop correction, including both bulk and boundary
contributions. Note that, with $v$ being just the scale of EWSB, the
extent of mixing is miniscule even at $R^{-1} = 500$ GeV and is
progressively smaller for the higher KK-modes. As a consequence, unless 
$R^{-1}$ is very small, the  $Z^{1}$ and $\gamma^1$ are, 
for all practical purposes, essentially 
$W^1_3$ and $B^1$. This has profound
consequences in the decays of the KK-excitations.

With the mass corrections for $B^n$ always being negative, 
and the contribution from EWSB being relatively small, it is 
understandable that the KK-photons have masses very close to $n/R$. 
A further consequence is that the lightest KK-particle (LKP), and 
hence the Dark Matter candidate, 
is almost always the $\gamma^1$. Furthermore, of the other KK-excitations, 
the leptonic fields are the lightest (and nearly degenerate) followed by the 
weak gauge boson excitations. The quark and gluon excitations are the 
heaviest within a given KK-level on account of the substantial corrections 
due to the strong interactions. 

In a similar vein, the quark excitations mix too. With the
5-dimensional theory necessarily being a non-chiral one, for every SM
fermion state, there correspond two states at each KK-level (see
eqn.\ref{mode_expan}). These obviously mix with each other with a strength 
determined by the SM Yukawa coupling. Thus, while the two top-states at 
level one (hereafter denoted as $t^1_1$ and  $t^1_2$) are split slightly, 
the other KK fermions are almost degenerate. 
This is depicted in Table \ref{kkmass} which lists 
the masses of the first KK-modes of various
SM particles, after the inclusion of radiative corrections, 
 for three representative valuesof $1/R$, namely, 300, 500 and 800 GeV.

\begin{table}[tbh]
\begin{center}
\begin{tabular}{|c|c|c|c|c|c|c|c|c|c|}  \hline 
$R^{-1}$ &$m_{t^1_1}$ & $m_{t^1_2}$ & $m_{b^1_1},m_{b^1_2} $ & $m_{L^1}$ & $m_{e^1}$ & 
$m_{g^1}$ & $m_{{W^{\pm}}^1}$ & $m_{Z^1}$ & $m_{\gamma ^1}$ \\ \hline 
300 &389.0 & 373.5 & 347.6 & 309.0 & 303.3 & 384.3 & 327.4 & 328.6 & 301.5 \\ 
500 &599.9 & 587.3 & 585.5 & 515 & 505.5 & 642.3 & 536 & 542.1 & 501.0 \\ 
800 & 941.6& 900.3 & 924.9 & 815.9 & 808.7 & 1024.7 & 850.4 & 850.5 & 800.0 \\ \hline 

\end{tabular}
\end{center}
 \caption{{\it One-loop corrected masses (in GeVs) for the 
 first KK-mode of different fermions and gauge bosons for 
three values of the compactification radius $R$.}}
\label{kkmass}
\end{table}

\section{Production and decay of the first level KK-top and bottom quarks}
\label{production}

The KK-quarks being strongly interacting, their pair
production at the LHC is suppressed only by their large mass.
As can be easily appreciated, for $R^{-1} \gtap \, 500 \gev$, the 
KK-excitations of different flavors are close-spaced in mass, leading 
to near-democratic production cross 
sections\footnote{The democracy, of course, is not strictly true, as the 
production of first family quark excitations and/or gluon excitations
(in various combinations) receive substantial $t$-channel contributions.}. 
The KK-top and KK-bottom quarks are somewhat special though, 
as they produce
$b$-quark jets in their decays.
Since these can be identified with a high efficiency, it 
makes their detection easier
than that of the excitations of the first two generations.

Pair production of 3rd generation KK-quarks in a proton (anti-)proton
collision proceeds in a fashion largely analogous to that applicable
to SM heavy quarks, the analytic expressions for which can be found in
Ref.\cite{Barger-Phillips}.  An extra contribution appears though,
namely in the form of a $s$-channel exchange of a level--2 KK
gluon\footnote{Note that the coupling of a pair of SM gluons to the
level-2 gluons violates KK-number but not KK-parity. Consequently, it
is allowed, but is loop-suppressed.}.  Since the level-2 gluon is
allowed to be (nearly) on-shell, this contribution is not negligible,
and indeed a careful calculation shows that, numerically, its effect
can be as large as 10\%.

\begin{figure}[htb] \begin{center}
\includegraphics[width=8cm,height=9.0cm,angle=-90]{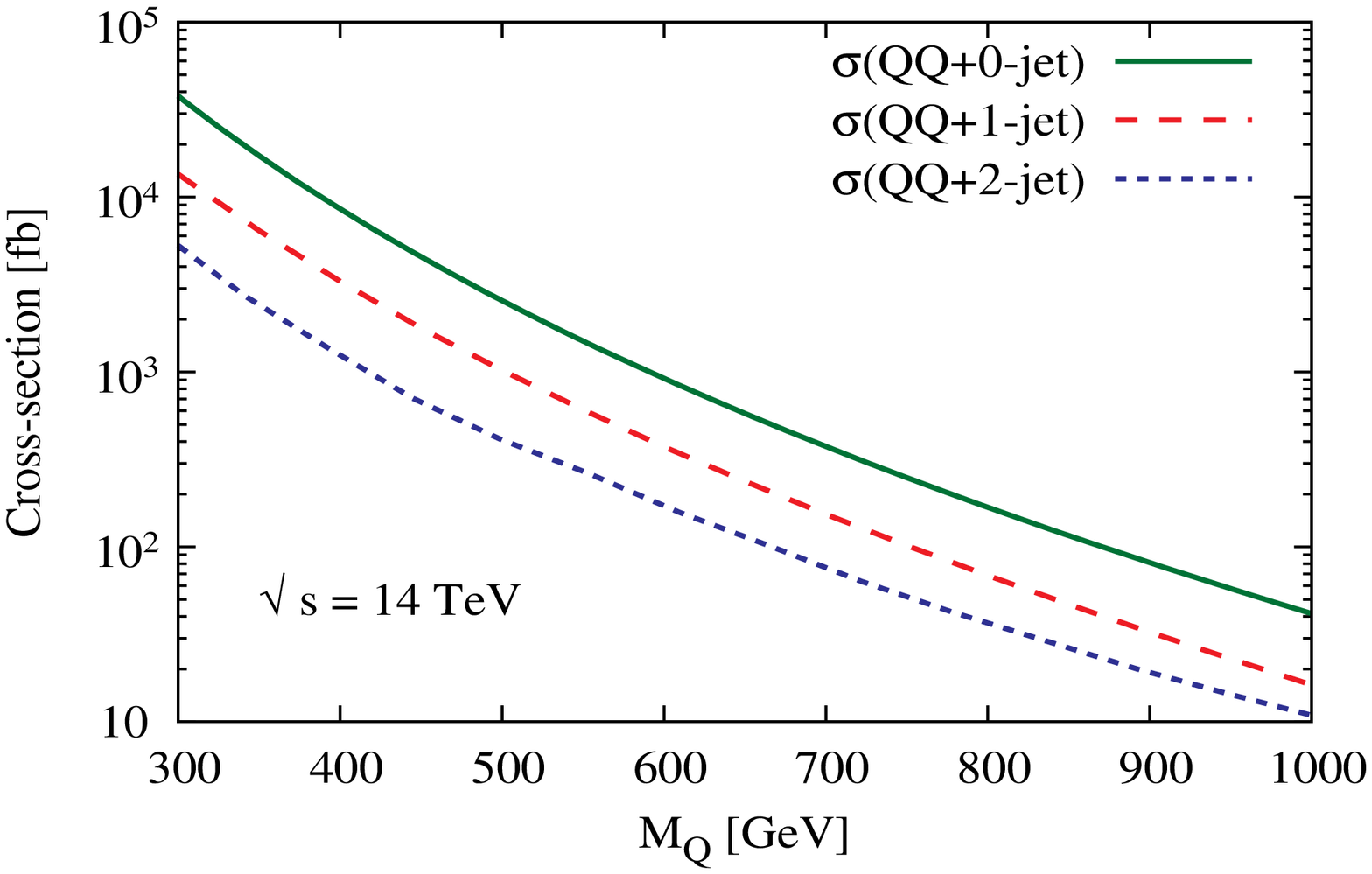}
\hspace*{-2cm}
\includegraphics[width=8cm,height=9.0cm,angle=-90]{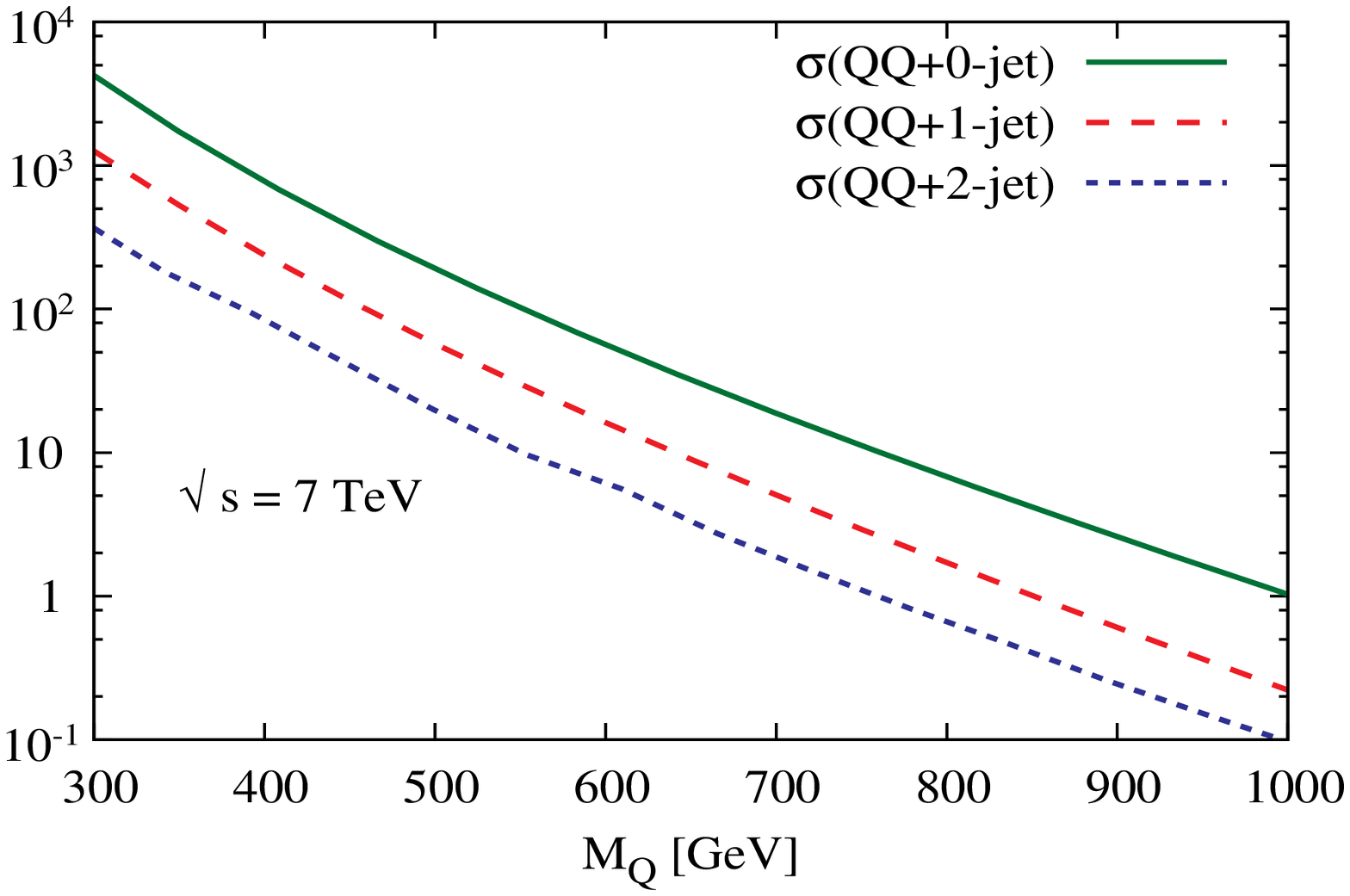}
\caption{\it Pair production cross-sections of level-one KK $t$-- or
$b$--quarks as a function of their masses
in proton proton collisions at center of mass energies 14 TeV
and 7 TeV respectively. 
Also shown are the cross-sections 
for pair-production in association with one or two hard jets (see eq.\ref{def:ratio}).}
\label{fig:cross-sect}
 \end{center} \end{figure}

For numerical evaluation of the cross-sections, we use a tree-level
Monte-Carlo programme incorporating CTEQ6L \cite{cteq6} parton
distribution functions. Both the renormalization and the factorisation
scales have been set equal to the subprocess center-of-mass energy
$\sqrt{\hat s}$.  The ensuing cross-sections are
presented in Fig.\ref{fig:cross-sect} for two different values of the
proton-proton center of mass energy.  Given the typical spectrum (displayed 
in Table \ref{kkmass}), it is immediately apparent that 
the cross-sections for $b^1_1$-pair production would be essentially 
identical to those for $b^1_2$, while $t^1_2$-pair production rates 
would slightly exceed those for  $t^1_1$. 
While the NLO and NLL
corrections can be well estimated by a proper rescaling of the
corresponding results for $t \bar t$ production, we deliberately
desist from doing so. With the $K$-factor expected to be
large~\cite{ttbar}, our results would thus be a conservative one.

At this stage, it is worthwhile to consider the possibility 
that the pair-production be accompanied by other hard jets. This 
would be of particular importance in estimating the background from
$ t \bar t$--production as discussed in the next section. Note that,
nominally, the production of extra light jets is associated with 
soft and collinear singularities. Parts of such processes are, 
of course, absorbed in defining the higher-order (NLO/NNLO, as the case 
may be) cross-sections. 
Defining, somewhat arbitrarily,
\beq
\sigma(Q \bar Q + n_j\mbox{--jets}) 
\equiv \sigma[p_T(j) > 30 \gev, \, 
                                |\eta(j)| < 2.5, \,  \Delta R(j-j) > 0.7]   
\label{def:ratio}
\eeq
where the last requirement is applicable for $n_j \geq 2$ and serves to
eliminate a particular collinear singularity by demanding a minimum
distance among them in the $\eta - \phi$ plane, with $\phi$ being the
azimuthal angle, and $\Delta R \equiv \sqrt{(\Delta \eta)^2 + (\Delta
\phi)^2} $.  In Fig.\ref{fig:cross-sect}, we also plot these
cross-sections $n_j = 1, 2$. For $n_j \ge 3$, the corresponding 
cross sections are too small to be of relevance. 
Note that the ratios $\sigma(n_j = 1,2) / \sigma(n_j = 0)$ 
actually increases marginally with $M_Q$. It should be realised that the simultaneous use of eq.\ref{def:ratio} alongwith $K$-factors could result in some double-counting, and, ideally one should employ higher-order monte-carlo generators. However, given the kinematic requirements we impose, the extent of this error is miniscule and this does not alter the results in any significant way.  

Let us now turn our attention to the decay patterns for each of  these 
KK-excitations. $t_1^1$ and $b_1^1$ being dominantly
right-handed, have very small decay widths into $W^1$ and $Z^1$.
Consequently, the $b_1^1$ decays into a $b-\gamma^1$ pair 
with almost a 100\%
branching ratio.  However, $t^1_1 \rightarrow t \gamma^1$ is
kinematically disallowed unless $R^{-1} > 1 \tev$. For smaller
$R^{-1}$, the only allowed decay mode for the $t^1_1$ is into $H^{1+} \; b$. 
Similarly, kinematics dictates that 
$t^1_2$ decays dominantly to $W^1 b$ and $b^1_2$ decays to $Z^1 b$
in spite of the fact that either of these KK-quarks
couple to both $W^1$ and $Z^1$.

While $\gamma^1$ is the LKP and hence stable (escapes detection), 
both $W^1$ and $Z^1$
must decay. The $Z^1$ is lighter than the excited
quark states. Furthermore, there is no tree-level 
$Z^1 Z \gamma^1$ coupling. 
Thus, the dominant decay modes of the $Z^1$ are into the 
$(L^1)^\pm L^\mp$ and $\nu^1 \nu$ final states with nearly equal branching
ratios\footnote{The decay  $Z^1 \rightarrow (l^1)^\pm l^\mp$ is
suppressed by $\sin^2\theta^1_W$.}.  In the former case, the 
excited lepton decays subsequently ($L^1 \rightarrow L \gamma^1$).
This leads to a $Z^1$ signal consisting of two (oppositely)
charged leptons and missing energy.
Kinematics does not permit the
$W^{1\pm}$ to decay either hadronically or in the $W^\pm \gamma^1$ channel.
Consequently, it decays to either $L^\pm \nu^1$ or to $L^{1\pm}
\nu$ with equal branching ratios. The final decay products are
therefore $L^\pm \nu \gamma^1$ of which only the charged leptons are
observable. Hence, we have
\beq
\barr{rclcl}
t^1_2 & \to & W^{1+} + b & \to & b + L^+ + \nu + \gamma^1 
\\[2ex]
b^1_2 & \to & Z^1 + b & \to & b + \nu^1 + \nu, \ b + L^+ + L^- + \gamma^1
\earr
\eeq
Thus, $t^1_2$-pair production would dominantly lead to a pair of $b$-jets, 
a pair of opposite sign-leptons (not necessarily of the same flavor) 
and missing transverse energy. The same configuration is also 
reached in nearly half the events resulting from $b^1_2$-pair production. 
The other half of the latter case is shared roughly equally between 
($ 2 \, b + p_T\hspace*{-1em}/\hspace*{1ex}$) and 
($ 2 \, b + 4 \, \ell + p_T\hspace*{-1em}/\hspace*{1ex}$) 
events\footnote{In this discussion, we, obviously, have 
confined ourselves to the $n_j = 0$ sector above. The inclusion of 
$n_j \neq 0$ events would lead to additional (non-$b$) jets.}.

Let us also briefly comment on the possible decay of the $H^{1\pm}$,
a by-product of $t^1_1$ production.  As the level-1 Higgs masses may
receive a substantial contribution from the boundary Higgs mass 
$\overline m_H$ (see eqn.\ref{delta1}), their decay patterns are crucially
dependent on this quantity. For simplicity, we have set $\overline m_H
= 0$ in our analysis. Given this, kinematical constraints allow only
$H^{1-} \to \tau^{1}_1 + \bar \nu_{\tau}$.  The tiny mass
separation between the LKP and $\tau^{1}_1$ ensures that the latter
subsequently decays to a $\gamma^1$ and a very soft $\tau$.  Such soft
$\tau$'s, whether from this decay chain or produced otherwise (such as
from the decays of $\tau^{1}_2$, which are produced from $W^{1\pm}$
or $Z^{1}$) often escape detection in a hadronic environment. So in
the following analysis we will only bank on the detection of prompt
$e$'s and $\mu$'s. Of course, the inclusion of a non-zero $\overline m_H$
would open other decay channels to the $H^{1\pm}$, thereby raising the 
possibility that some of the $t^1_1 \bar t^1_1$ events would lead to
final states observable above the SM background. This, however, would 
only add to the significance of our numerical results, at the cost of 
including an additional parameter. Once again, we desist from such 
a course of action.

\section{Collider searches}

As we have argued above, the dominant by-products of $t^1_2 \, \bar t^1_2$ 
and $b^1_2 \, \bar b^1_2$ production are a pair of $b$-quarks, a pair of
oppositely charged leptons (not necessarily of the same flavor) 
accompanied by missing transverse energy. This, of course, constitutes a 
classic top-quark detection strategy. It, thus, might be tempting to 
search for such states at the Tevatron. However, for $R^{-1}$ allowed 
currently, the production rates at the Tevatron are too small to be 
of any relevance. Hence, we concentrate on the LHC.

With the relative mass splitting between the excited quarks and the
weak gauge bosons ($Z^1$ and $W^1$) and, in turn, the mass splitting
between the gauge bosons and the LKP being small, it is quite obvious
that, in the rest frame of the produced excited quark, the daughters
$b$ and the lepton would carry only a small fraction of its mass as
momenta.  This, in turn, translates to relatively small transverse
momenta for them even in the laboratory frame. Thus, one should
profitably concentrate on such a part of the phase space. A fruitful
pursual of such a methodology requires that we examine and compare the
phase space distributions for signal events as well as those of the
myriad backgrounds which we discuss below.

There are several SM processes which result in final states similar to 
that we are interested in. Consequently, one has to find the
characteristics of our signal which are distinct from the SM
processes.  It is needless to mention that SM $t\bar t$ production (at
the LHC), being driven by strong interactions, offers us the
biggest hurdle. Apart from the $t\bar t$ production, non-resonant, $b
 \bar b W^+ W^-$ and $b \bar b Z Z$ production 
$(\rm{\cal O} \left(\alpha_s \alpha_W \right)
)$ followed by the leptonic decays of the $W (Z)$-bosons, give rise
to analogous final states. Drell-Yan production of leptons in association with
two light flavored jets, misidentified as $b$-jet may also contribute
to the background. In the following we will discuss all of these
one by one.

However, before we embark on the mission to suppress the above mentioned 
backgrounds, it behoves us to list the 
basic requirements for jets and isolated leptons to be visible as 
such. In this quest, it should be 
realized though that any detector has only a finite resolution. For a 
realistic detector, this applies to both energy/transverse momentum 
measurements as well as determination of the angle of motion. For 
our purpose, the latter can be safely neglected\footnote{The angular 
resolution is, generically, far superior to the energy/momentum resolutions
and too fine to be of any consequence at the level of sophistication of this
analysis.} and we simulate
the former by smearing the energy with 
Gaussian functions defined by an energy-dependent width. 
The latter receives corrections from many sources and
are, in general, a function of the detector coordinates. We, though, 
choose to simplify the task by assuming a flat resolution function 
equating it to the worst applicable for our range of interest. To wit, 
we have
\beq
  \frac{\sigma_E}{E} = \frac{a}{\sqrt{E}} \oplus b 
\eeq
where the errors are to be added in quadrature and 
\beq
\barr{rclcrclcl}
a_\ell &= & 0.05 \ , &\quad& b_\ell &= &5.5 \times 10^{-3} \ & \qquad &
 {\rm for ~leptons,}
\\[1ex]
a_j &= &0.80 \ , & &  b_j &= &0.05 & & {\rm for~ partons}.
\earr
\eeq

Keeping in mind the LHC environment as well as the detector
configurations, we demand that, to be visible, a jet or a lepton must
have an adequately large transverse momentum and they are well inside
the rapidity coverage of the detector, namely,
\beq
p_T({\rm jet}) > 30 \gev \ , \qquad p_T({\rm lept}) > 15 \gev \ ,
\label{cut:pT}
\eeq
and
\beq
|\eta({\rm jet})| \leq 2.5 \ , \qquad |\eta({\rm lept}) | \leq 2.5 \ .
\label{cut:eta}
\eeq
Furthermore, we demand the leptons and jets be well seperated from
each other by requiring 
\beq
\Delta R_{\ell -j} \leq 0.4 ~{\rm and}~ \Delta R_{j \, j} > 0.7 \ .
\label{cut:jj-iso}
\eeq

Among the SM backgrounds mentioned above, $t \bar t$ production
(followed by $t \rightarrow b W$ and $W \rightarrow l \nu$) is very
large, even on the imposition of eqs.(\ref{cut:pT}--\ref{cut:jj-iso}).
Another set of important backgrounds arise from inclusive $W$-- 
and $Z$--production.
To reduce this to manageable levels, one must impose further selection
cuts. Since the only hadronic constituent of our signal is given by a
pair of $b$'s, we require that there be two and only
two\footnote{$b$-tagging can also control the background from DY
production of leptons mentioned above.
}  
$b$-jets satisfying
eqs.(\ref{cut:pT}--\ref{cut:jj-iso}).
  For $b$-tagging efficiency, we
use $\epsilon_b = 0.6$ for each $b$. Furthermore, the event must be
characterized by a minimum missing transverse momentum defined in
terms of the total visible momentum, namely,
\beq
\not p_T \equiv \sqrt{ \bigg(\sum_{\rm vis.} p_x \bigg)^2 
                 + \bigg(\sum_{\rm vis.} p_y \bigg)^2 } \; > 30 \gev \ .
\label{cut:pt-mis}
\eeq 

While this seemingly restricts the potential SM background to 
$ t \bar t$ production, in reality, a multitude of 
other electroweak processes contribute too. 
For some of these, the hard (parton level) process does not even 
have a source of missing energy. For example, 
even a $2 \, b + 2 \ell$ final state could potentially 
be associated with a missing transverse momentum simply on account 
of mismeasurements of the jet and lepton energies. A minimum requirement of 
the missing transverse momentum keeps these backgrounds well under control.
The
requirements summarized by Eqns.(\ref{cut:pT}--\ref{cut:pt-mis}),
thus, constitute our {\it acceptance cuts}.

A further requirement proves useful. Most such electroweak 
processes involve production of $W$ or $Z$ bosons,
and these could be decimated by the simple requirement of the dijet 
and dilepton (in the identical-flavor case) invariant masses 
to be substantially away from these masses. This leads to the 
first of our selection cuts, namely
\beq
  m{(\ell, \ell)} \not \in [70, 100] \gev 
       \label{cut:invmass}
\eeq
being applicable only to the case of final states with the oppositely
charged hard leptons being of identical flavor. While it might seem
appropriate to impose a similar restriction on the dijet invariant
mass, we desist from doing so. For one, once we require $b$-tagging
for the two hardest jets, the dominant ($t \bar t$) background is not
characterized by any such peak in $m_{b \bar b}$.  As for the
subdominant background sources wherein the jets are the daughters of a
$Z$, the other cuts suffice. And, finally, such a requirement
adversely affects the signal strength as we shall see below.

Backgrounds such as inclusive $W / Z$ production (single or multiple
gauge bosons) were estimated with the generator {\sc
Alpgen}\cite{Mangano:2002ea} and found to be in consonance with those
obtained by the ATLAS collaboration after an exhaustive detector
simulation\cite{Aad:2009wy}. For example, after imposition of the
above cuts, the contributions to the background cross-section from
non-resonant $bbWW$ production (and $W$-decay) is 65 fb, whereas the
contributions from $bbZZ$ and $WWjj$ production (wherein two light
flavored jets are misidentified as $b$-jets\footnote{A mistagging
probability of 3\% has been assumed here \cite{bmistag}.}) are both
less than a fb.  In fact, one can completely get rid of $bbZZ$ events
by a cut on $m_{\ell \ell}$ (Eqn.\ref{cut:invmass}). Similarly, the
Drell-Yan production of leptons in association with two jets
(mis-identified as $b$-jets) can be reduced to virtually zero by the
cuts on the transverse momentum of jets (Eqn.\ref{cut:pT}), missing
transverse momentum (Eqn.\ref{cut:pt-mis}) and a cut on the invariant
mass of the leptons (Eqn.\ref{cut:invmass}). So contributions from
most of the processes listed earlier in this section have been reduced
to innocuous levels leaving SM $t \bar t$ production to dominate
overwhelmingly (after imposition of the acceptance cuts this
contribution is to the tune of 4.1 pb).

The latter was simulated at parton level using only the tree-level 
matrix elements
and incorporating CTEQ6L parton
distributions.  However, to take into account the higher
order corrections, we multiplied the rate by the appropriate
$K$-factor~\cite{ttbar}. 
For the subdominant backgrounds, whether 
these be from $t \bar t$ with additional hard jets (see eqn. \ref{def:ratio})
or inclusive $W/Z$-production,  we do not
include higher-order corrections.
The consequent error is miniscule.
For the signal though, we are faced with the non-availability of the
higher-order calculations. Given that the production is
QCD-driven, it is tempting to use the $t \bar t$ calculation and scale
it appropriately. However, we choose to desist from this. Considering
the fact that the $K$-factor for the signal events are expected to be
of the same order as those for $t \bar t$, our results thus are
conservative.

\begin{figure}[!htb]
\begin{center}
\vspace*{-2ex}
\includegraphics[width=6.0cm,height=5.5cm,angle=-90]{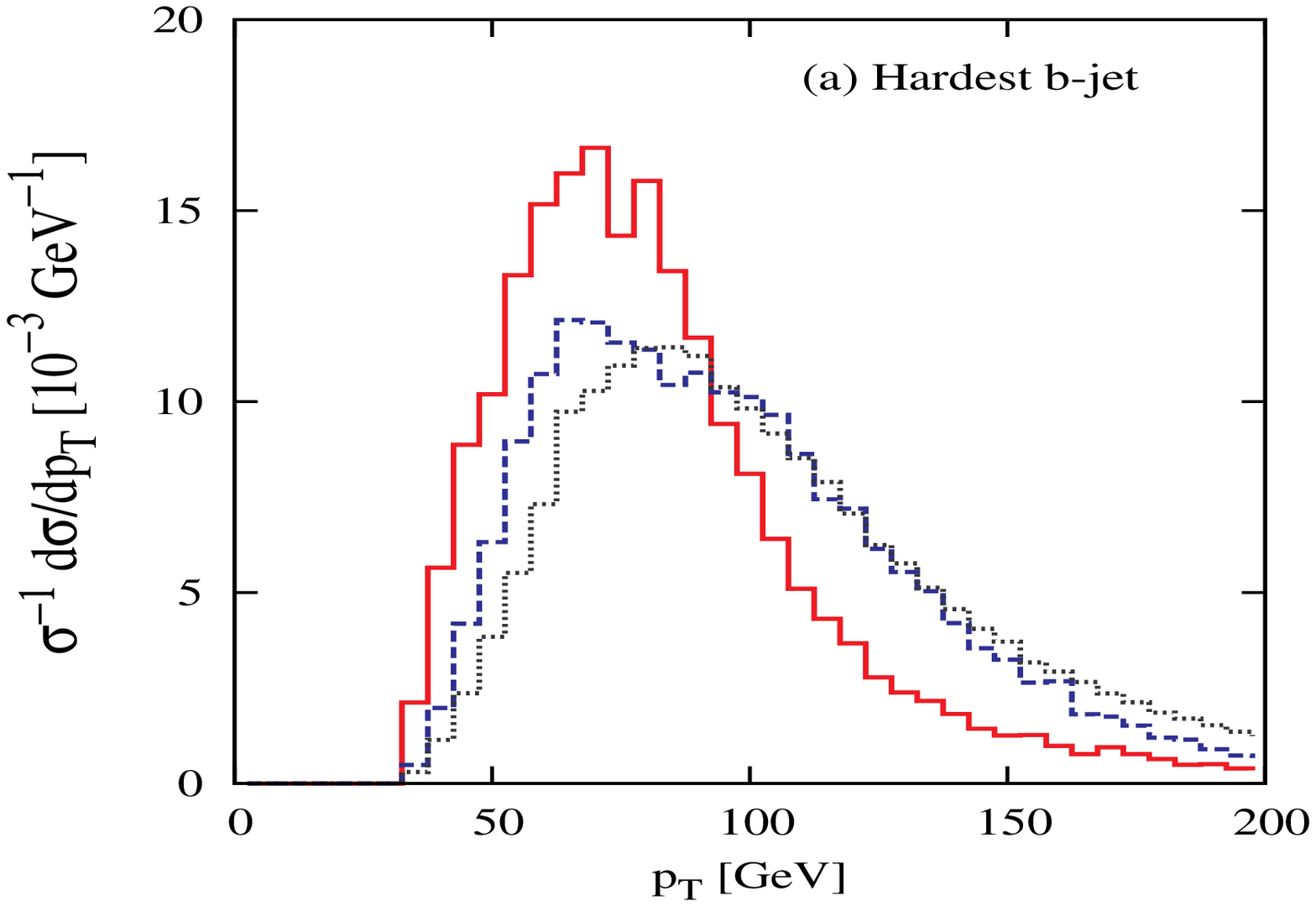}
\hspace*{-2ex}
\includegraphics[width=6.0cm,height=5.5cm,angle=-90]{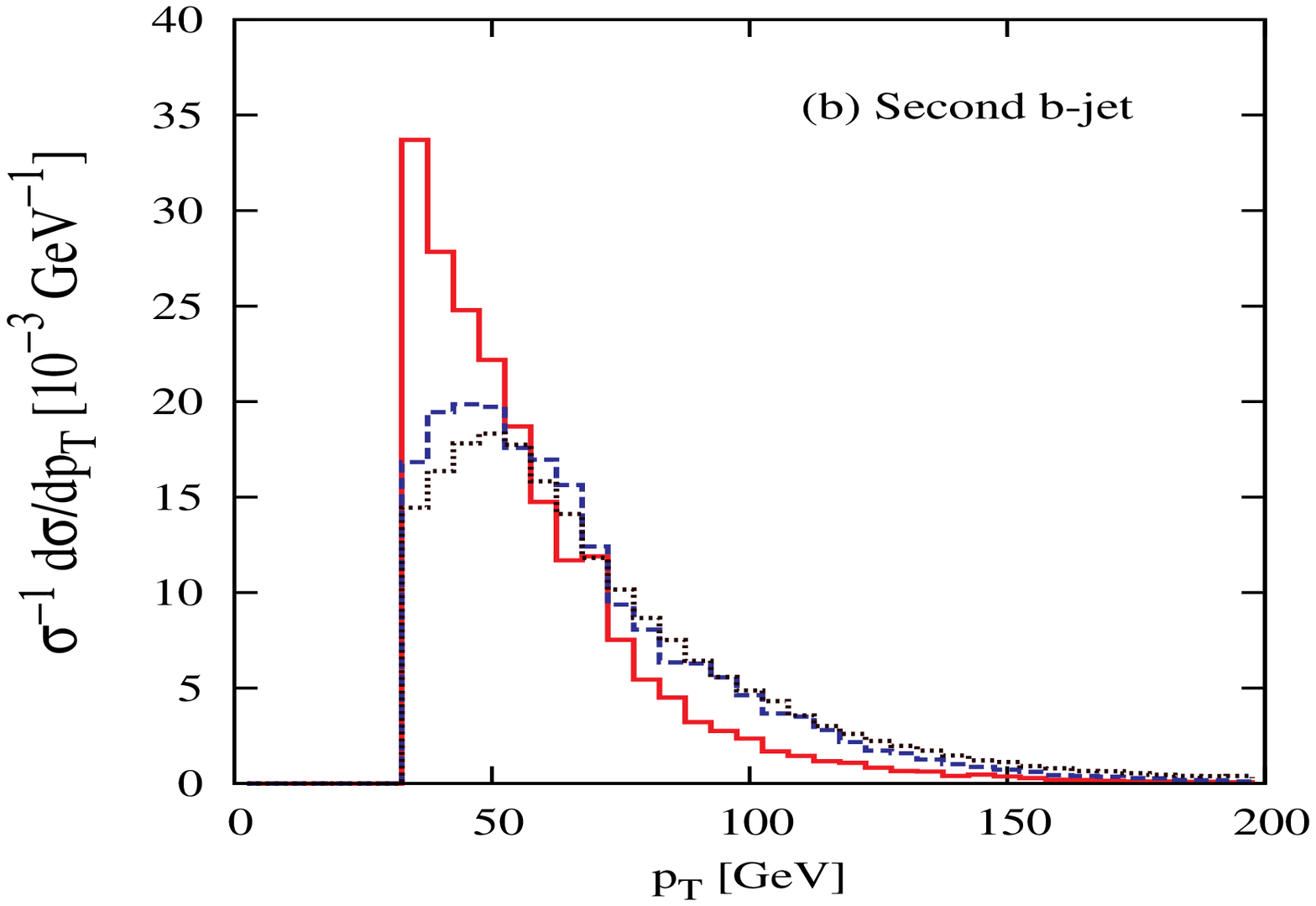}
\hspace*{-2ex}
\includegraphics[width=6.0cm,height=5.5cm,angle=-90]{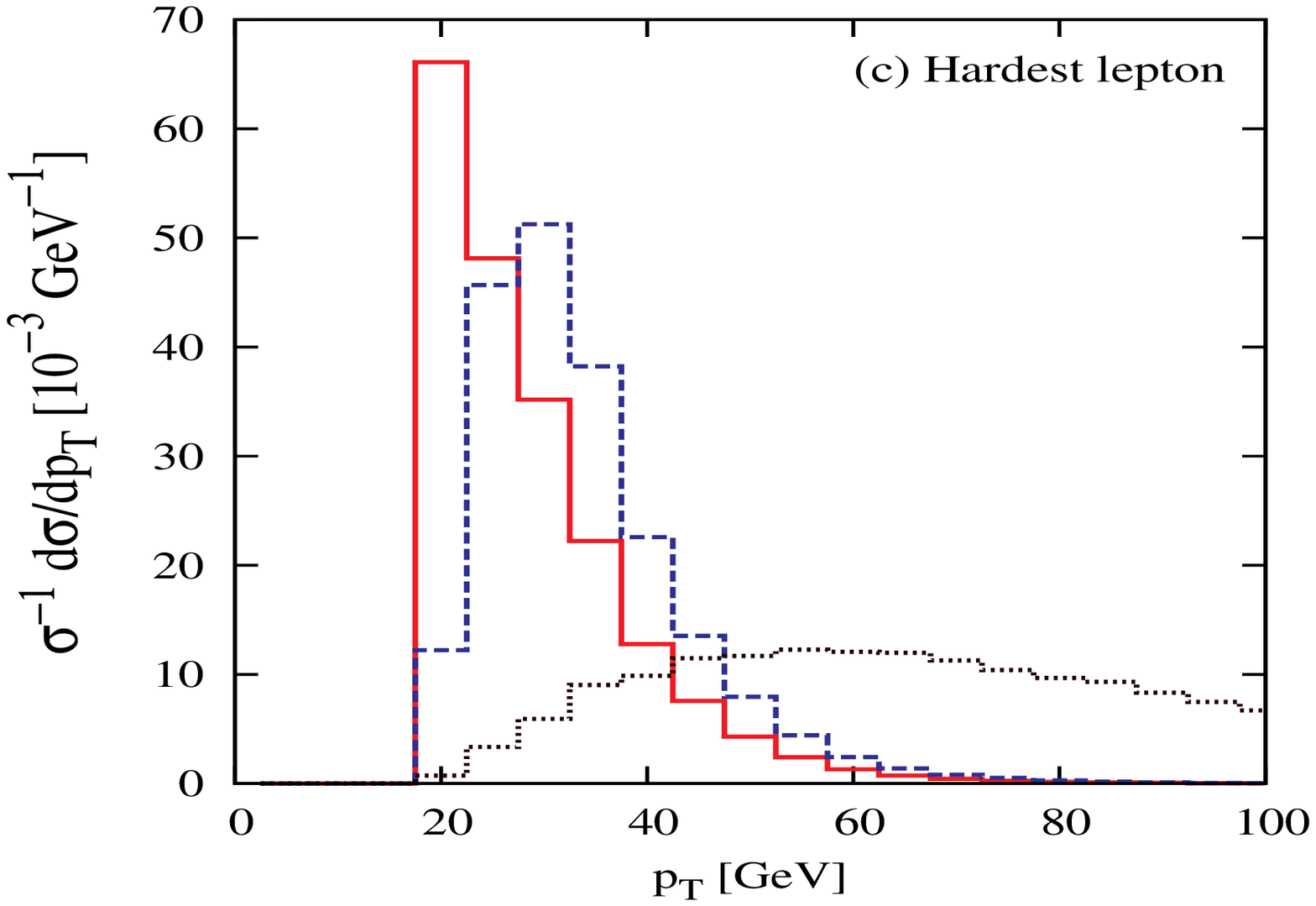}

\vspace*{-1ex}
\includegraphics[width=6.0cm,height=5.5cm,angle=-90]{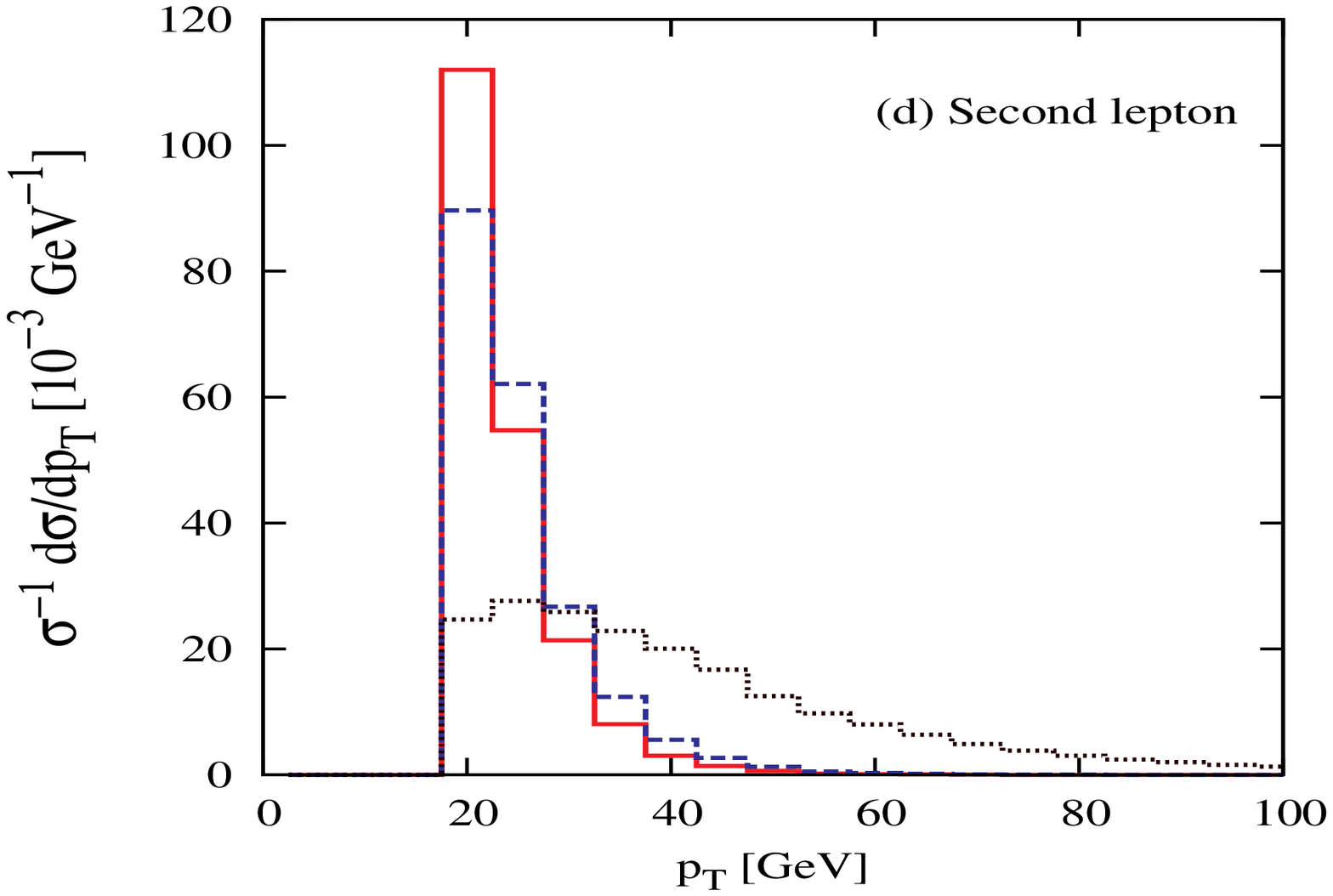}
\hspace*{-2ex}
\includegraphics[width=6.0cm,height=5.5cm,angle=-90]{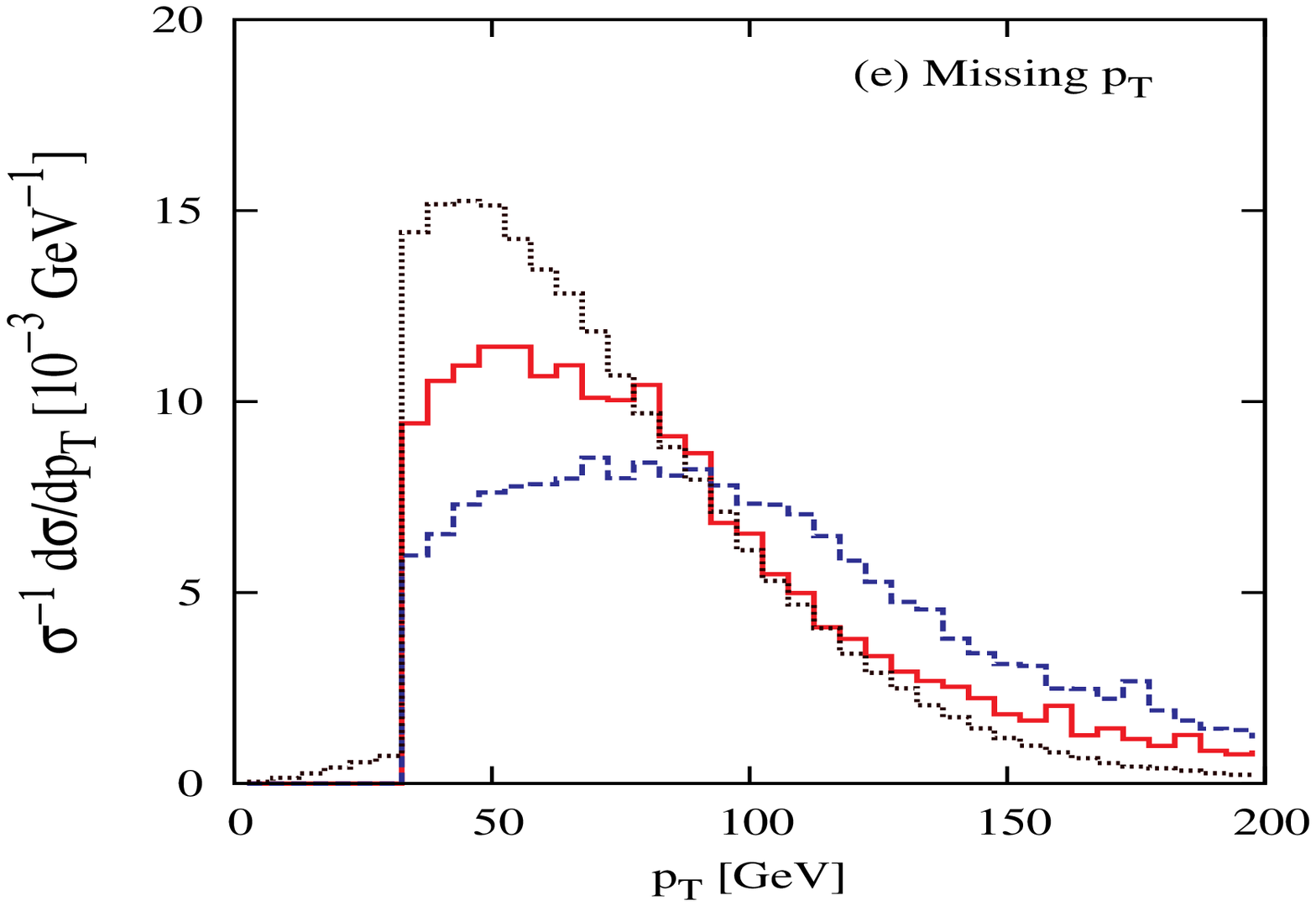}
\hspace*{-2ex}
\includegraphics[width=6.0cm,height=5.5cm,angle=-90]{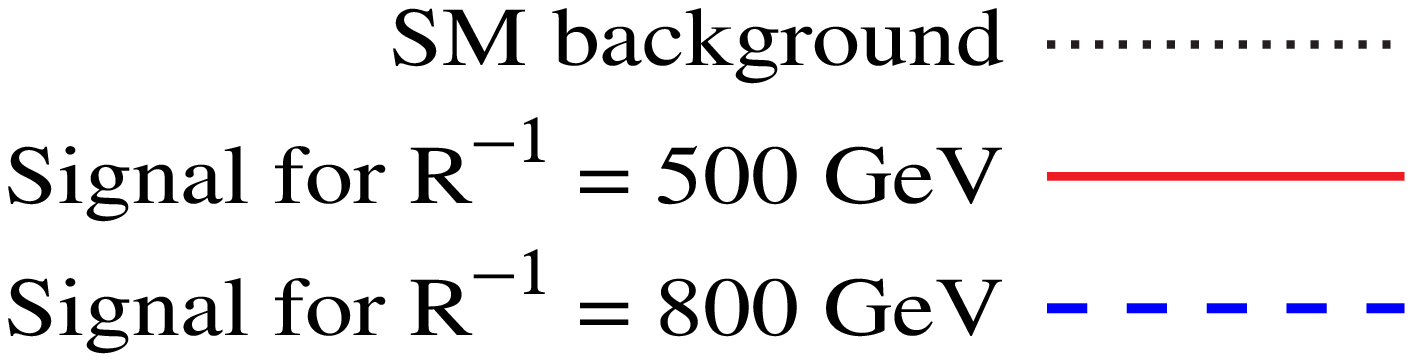}
\end{center}
\caption{{\it Normalized distributions for transverse energy on 
imposition of the acceptance cuts of Eqns.(\ref{cut:pT}--\ref{cut:pt-mis}).
{\it (a)} the leading $b$-jet, {\it (b)} the second $b$-jet, 
{\it (c)} the leading lepton, {\it (d)} the second lepton
and {\it (e)} the missing transverse energy.
Each panel shows
the distributions for two signal points as well as the total SM background.}}
   \label{fig:pTjet}
\end{figure}

Working at $\sqrt{s} = 14 \tev$ we present, in
Figs.\ref{fig:pTjet}({\it a, b}), the normalized $p_T$ distributions
for the two $b$-jets for signal events (two different values for $R^{-1}$
are used, namely 500 and 800 GeV) along with the total SM
background. As expected, the signal distributions are marginally
softer than the corresponding distributions for the background. 
This difference is less pronounced for larger $m_Q$ values though. 
However, even for $R^{-1} = 800 \gev$, the signal $p_T(b)$ distribution 
would have been softer than the background had we limited 
ourselves to only the 
exclusive $Q \bar Q$ and $t \bar t$ events (i.e., only $n_j = 0$ vide 
eq.\ref{def:ratio}).
 More
interesting---and typical to UED---are the transverse momentum
distributions for the two leptons (as displayed in
Figs.\ref{fig:pTjet}({\it c, d})). With the signal events being peaked
at a relatively small $p_T(\ell)$, this immediately suggests a way to
suppress the widely-distributed background by imposing an upper bound
on this variable, resulting in a marked improvement in the signal to
noise ratio. The missing--$p_T$ ($\not{p_T}$) spectra for the signal and background
are also presented in Fig.\ref{fig:pTjet}({\it e}).  In spite of the
fact that in UED a massive inivisible particle ($\gamma^1$) is
responsible for the missing $p_T$ (in comparison to the SM where it is
the neutrinos from various decay chains), the $\not{p_T} $ spectra do
not differ drastically.  However, it has been mentioned earlier that 
the cut on $\not{p_T} $ is very much
instrumental in reducing SM backgrounds (in particular QCD driven
ones) in which jet-energy/momentum mismeasurements give rise
to missing transverse momentum in the final state.

The relative softness of the leptons and jets is a generic feature of
the model and but a manifestation of the typically quasi-degenerate
mass spectrum of UED model. Note that these are the decay products of
a 1st KK-level state and that the conservation of KK-parity demands
that each such decay must have another such level-one KK excitation as
a daughter.  The near-degeneracy of the level-one KK-spectrum thus
predicts that the leptons and jets should be soft. 
Thus, apart from
the acceptance cuts, defined above, a restriction on the {\it maximum}
value of the lepton transverse momenta, would enhance the
signal to background ratio considerably. As can be expected, the 
transverse momenta of the final state particles are correlated. For example,
imposing such a restriction on the  lepton transverse momenta, also renders 
the $p_T(b)$ spectra for the signal to be softer than those for the 
background. Thus, a similar restriction on the {\it maximum}
value of the $b$-jet transverse momenta also serves to accentuate 
the signal.

\begin{figure}[t]
\begin{center}
\vspace*{-2ex}
\includegraphics[width=6.0cm,height=5.5cm,angle=-90]{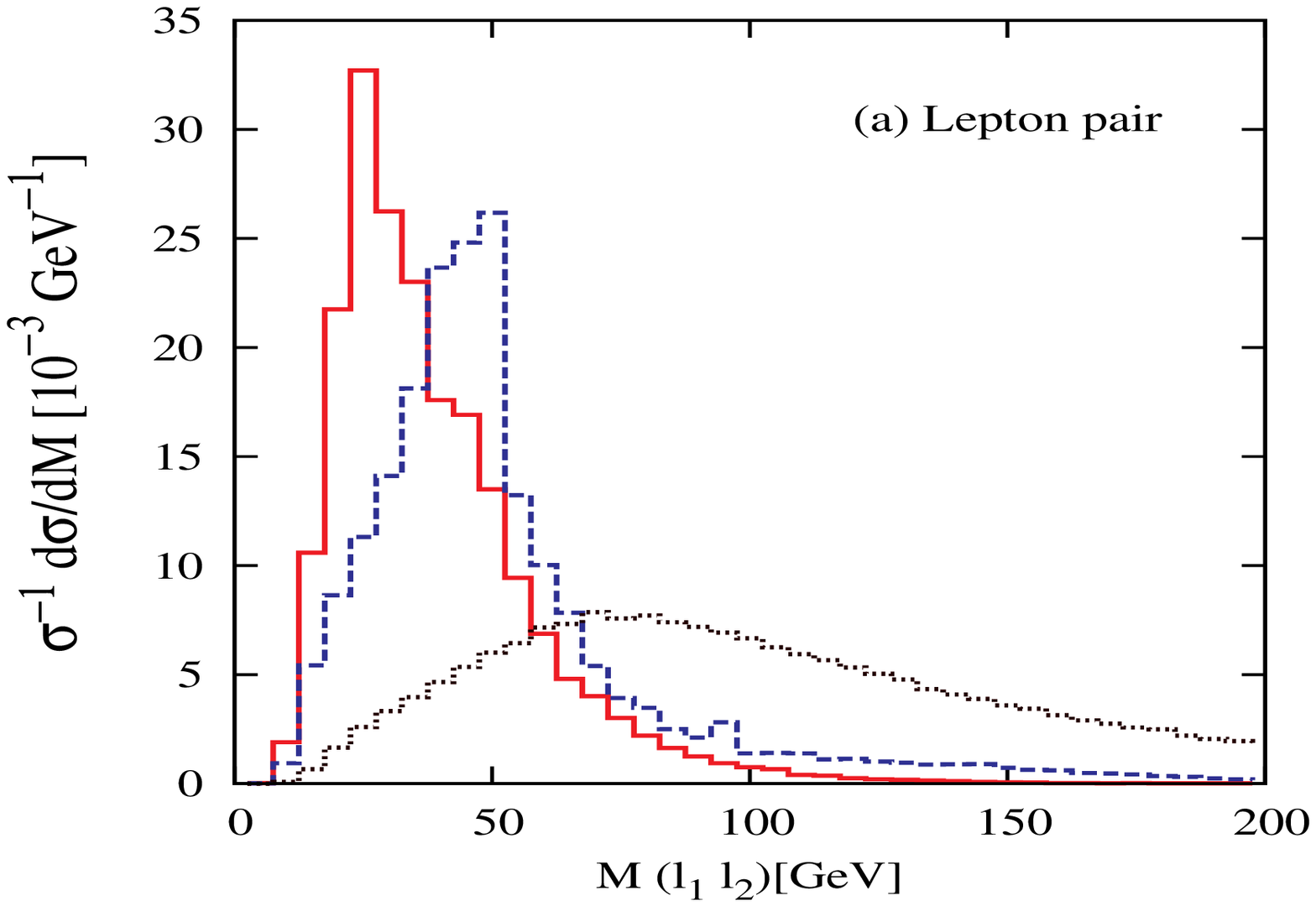}
\hspace*{-2ex}
\includegraphics[width=6.0cm,height=5.5cm,angle=-90]{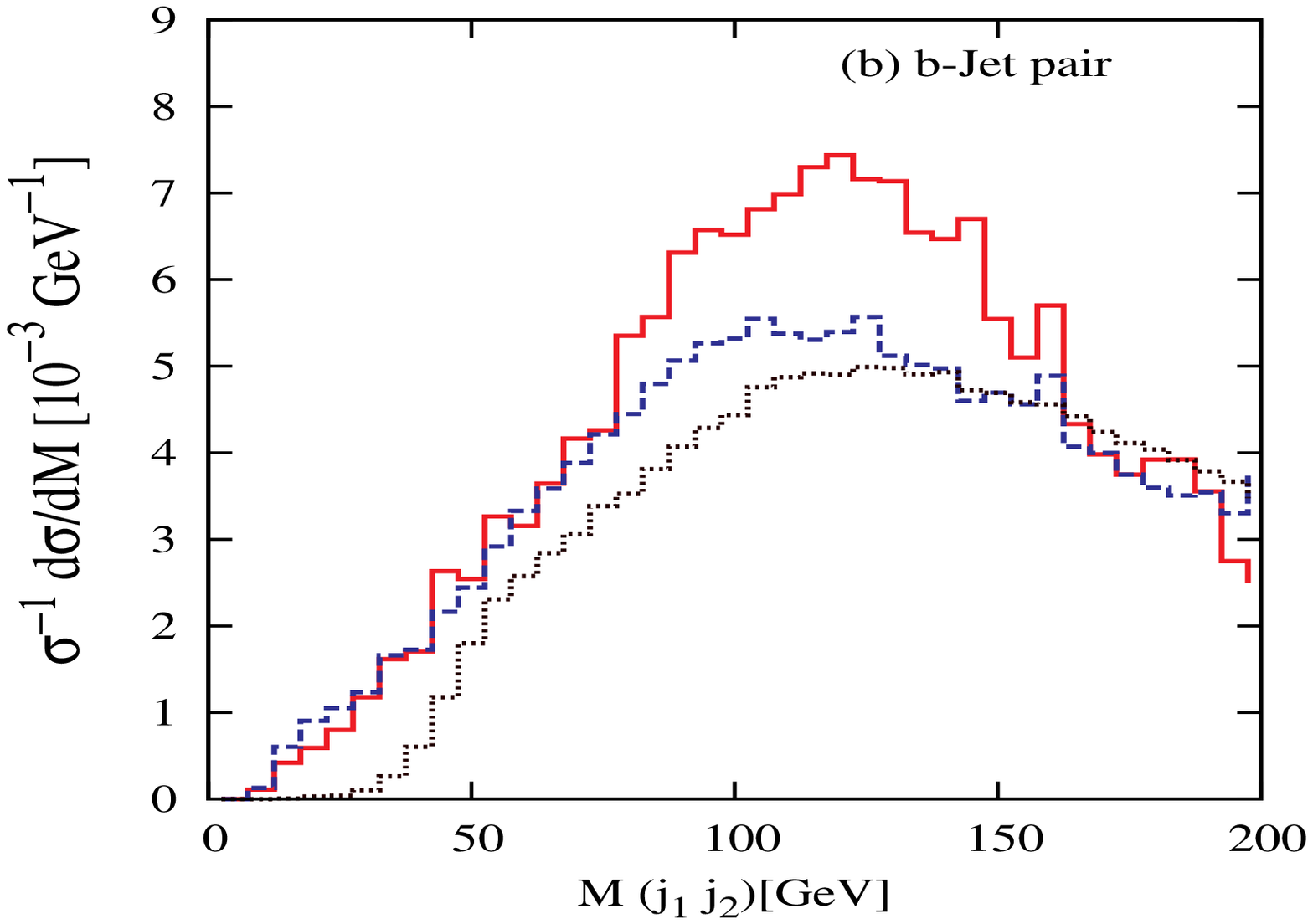}
\hspace*{-2ex}
\includegraphics[width=6.0cm,height=5.5cm,angle=-90]{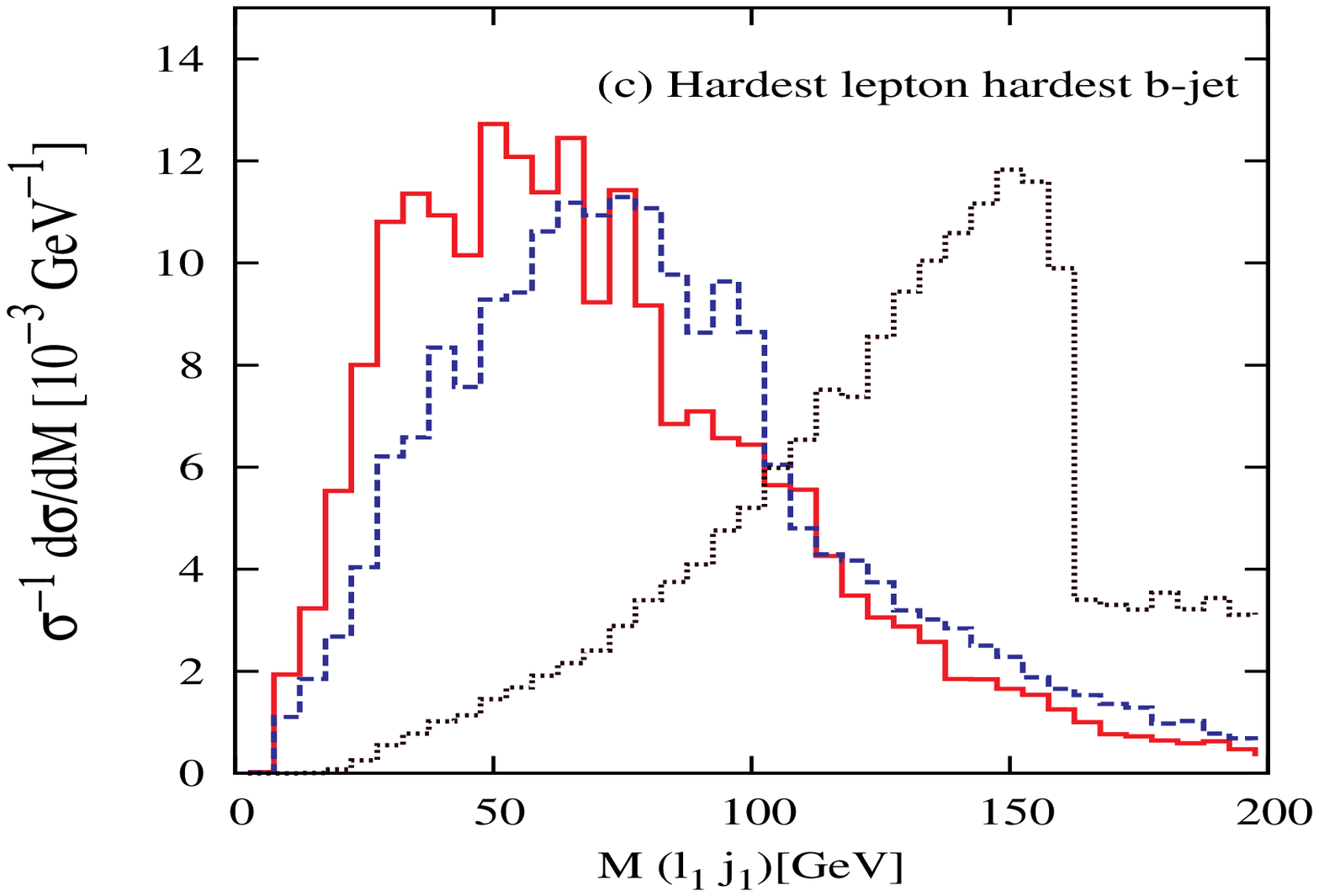}

\vspace*{-1ex}
\includegraphics[width=6.0cm,height=5.5cm,angle=-90]{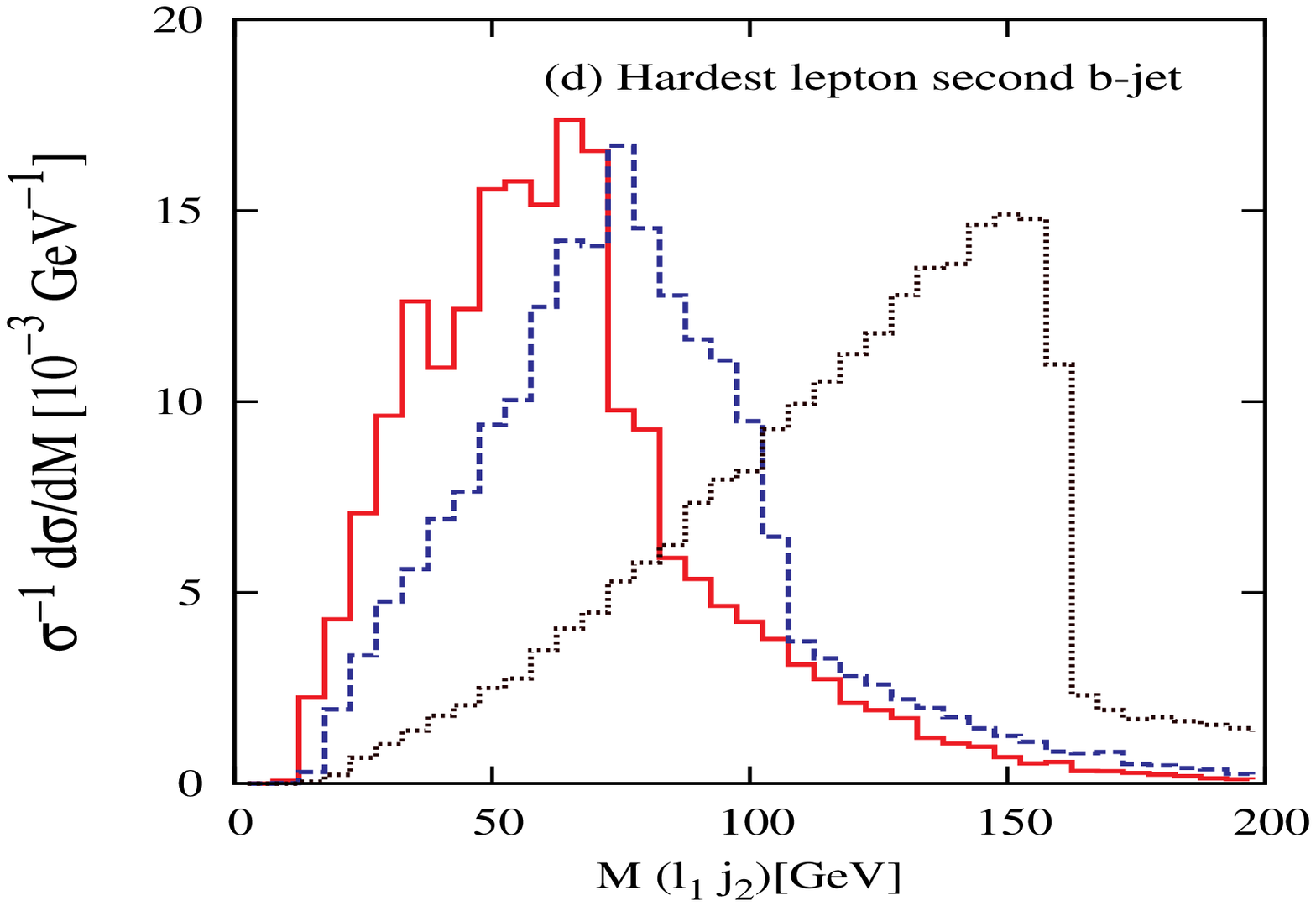}
\hspace*{-2ex}
\includegraphics[width=6.0cm,height=5.5cm,angle=-90]{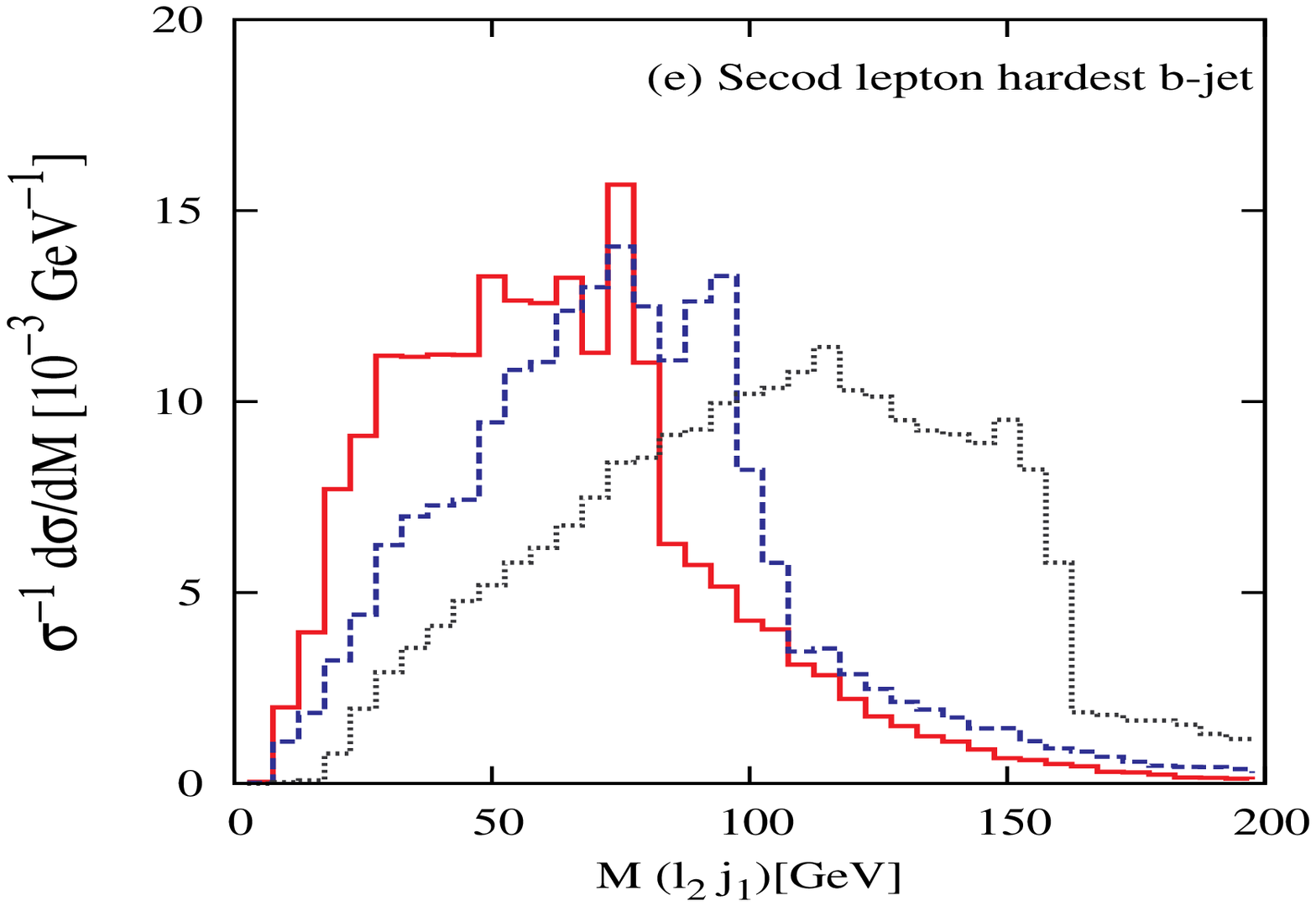}
\hspace*{-2ex}
\includegraphics[width=6.0cm,height=5.5cm,angle=-90]{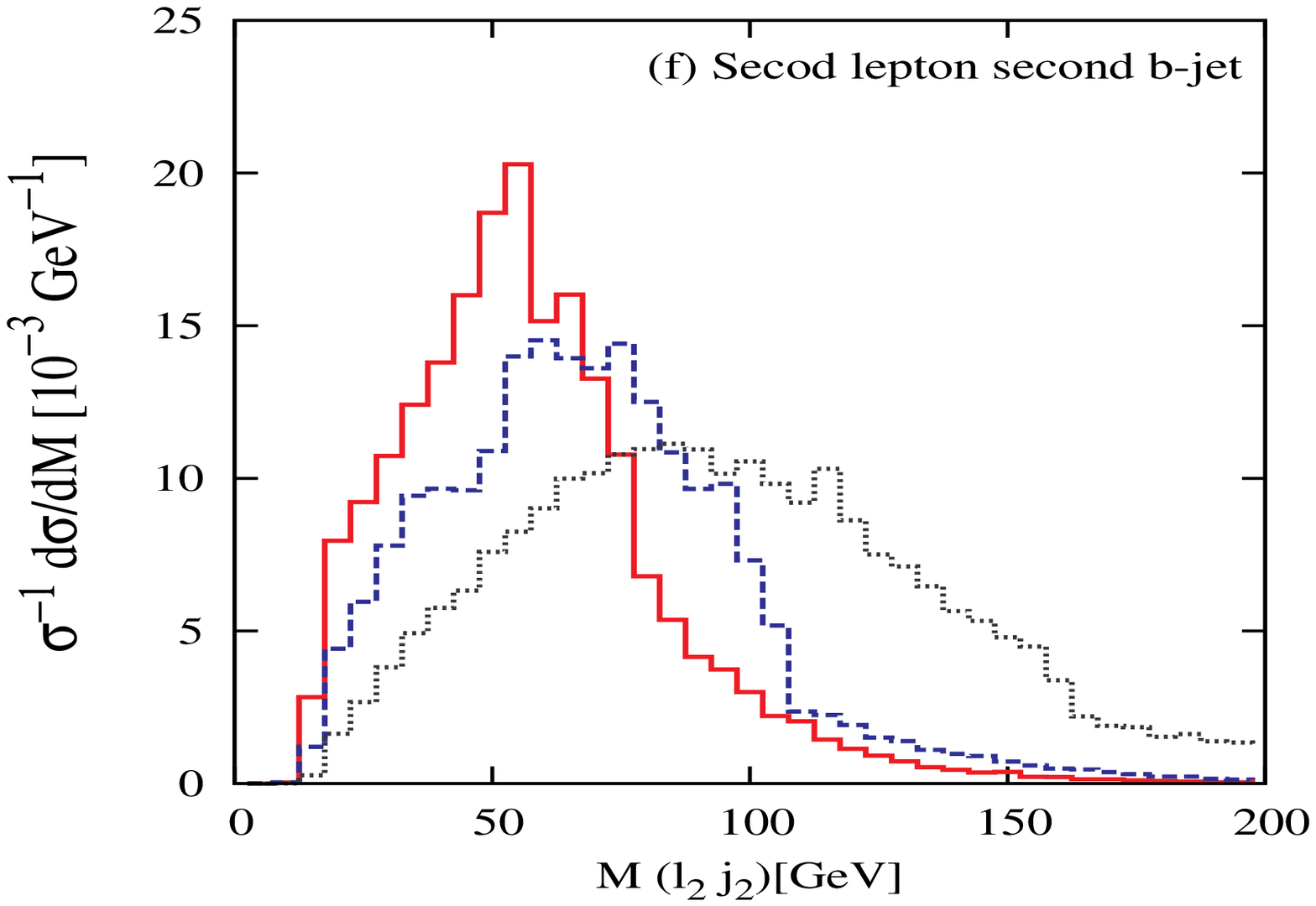}
\end{center}

\caption{{\it Normalized invariant mass distributions (on imposition of 
acceptance cuts only) for 
{\it (a)} lepton-pair, {\it (b)} $b$-jet-pair, 
{\it (c)} hardest lepton--hardest $b$-jet, 
{\it (d)} hardest lepton--second $b$-jet, 
{\it (e)} second lepton--hardest $b$-jet, 
{\it (f)} second lepton--second $b$-jet. 
In each case, the solid (red) and the dashed (blue) 
curves denote signal distributions for $R^{-1} = 500 (800) \gev$
respectively, while the dotted (black) curve denotes the 
total SM background.
}}
\label{invmass}
\end{figure} 

It is instructive to consider each of the six possible pairwise
invariant mass distributions constructed out of the momenta of the two
leading jets and the two leading leptons.  (see Fig.\ref{invmass}).
While charge measurement for the leptons is relatively
straightforward, that for the $b$'s has a relatively lower efficiency.
Rather than pay the price for this, we have identified both jets and
leptons by the relative magnitude of their transverse momenta. (The
lepton and $b$-jet with higher (lower) $p_T$ are denoted by $\ell_1$
($\ell_2$) and $b_1$ ($b_2$) respectively.) As Fig.\ref{invmass}$a$
shows, $M(\ell_1, \ell_2)$ peaks at relatively small values for the
signal events, while it has a flat distribution for the
background\footnote{While it may seem that the restriction of
Eqn.(\ref{cut:invmass}) is irrelevant, it should be realized that the
absence of this cut would have led to a an enhancement in the
background close to $M(\ell_1, \ell_2) \sim m_Z$ on account of
inclusive $W/Z$ (single or multiple) production processes.  That the
$M{(\ell_1, \ell_2)}$ distribution does not vanish exactly in this
regime is not surprising, for we do retain such events for
unlike-flavour lepton pairs.}. It is, thus, useful to concentrate on low
$M{(\ell_1, \ell_2)}$ events.  The $b$--$b$ invariant mass
distribution (Fig.\ref{invmass}$b$), on the other hand, is not an
efficient discriminator between signal and background. Indeed, had we
imposed a condition analogous to Eqn.\ref{cut:invmass} for this pair, 
we would have lost a larger fraction of the signal than the
background.

As for the various lepton--jet combinations, the sharp drop 
in the SM
background at $M{(\ell, b)} = m_t$ is a kinematic effect and 
characteristic of
the masslessness of the neutrino\footnote{The lack of such a drop in
Fig.\ref{invmass}$(f)$ is but a consequence of the fact that this
pairing is often the `wrong' one, namely the $b$-jet and the lepton
emanate from different top-quarks.}.  The corresponding drops in the
signal profiles occur at either of 
\[
M(\ell, b) = \frac{\sqrt{(m_{b^1 _2} ^2
- m_{Z^1} ^2) (m_{Z^1} ^2 - m_{L^1} ^2)}}{m_{Z^1}}
\qquad
{\rm and}
\qquad
M(\ell, b) = \frac{\sqrt{(m_{t^1 _2} ^2 - m_{W^1} ^2) (m_{W^1} ^2 -
m_{\nu^1} ^2)}}{m_{W^1}}
\]
depending on the particular production and decay channel involved. 
Given the near degeneracy of the masses, a resolution of the different
steps would require a very large luminosity, and for all practical purposes,
they essentially coincide. A similar feature is present in the dilepton
invariant mass distribution as well and the drop occurs at
\[
M(\ell_1, \ell_2) = \frac{\sqrt{(m_{Z^1} ^2 - m_{L^1} ^2) (m_{L^1} ^2 - 
m_{\gamma ^1}^2)}}{m_{L^1}} \ .
\]
On the
other hand, the $bb$ mass distribution does not show any such sharp
edge (either for the signal or background) as each of the jets are 
decay products of two different heavy quarks, thus carrying no
correlation in their mass distributions. 
As the structure of the bulk as well as orbifold 
corrections---eqs.(\ref{delta0}, \ref{delta1})---show,
the relative splittings between the KK masses is always
small. Thus, for the entire range of $R^{-1}$ likely to be accessible at the
LHC, this drop in the signal rates would occur well to the left of the 
top mass. This feature, then, can be used to define further 
selection cuts in the form of an upper bound on such jet-lepton 
invariant masses, especially for the combinations of Figs.\ref{invmass}.

Looking at the distributions in Figs. \ref{fig:pTjet}, \ref{invmass} and
following the above discussion, we are now in a position to propose
a further set of cuts, which, along with Eqn.(\ref{cut:invmass}) constitute 
our selection cuts, namely
\beq
\barr{rcl c rcl}
p_T^l & <  & 50 \gev 
\\[1ex]
p_T ^{b_1} & < &  100 \gev 
& \qquad & 
p_T ^{b_2} & < &   65 \gev
\\[1ex]
M(\ell_1,  \ell_2) & < &  70 \gev
& \qquad & 
M(\ell_1, b_1) & < &  100 \gev
\\[1ex]
M(\ell_1, b_2), M(\ell_2, b_1), M(\ell_2, b_2)  & < & 75 \gev
\earr
\eeq

\begin{table}[h]
\begin{center}
\begin{tabular}{||c||c|c|c|c|c||c||}
\hline \hline
 & \multicolumn{5}{|c|}{Cross-section in fb after cuts} & Total\\\cline{2-6}
 & &  &  & & $M(l_2,b_1)$, & after\\
& Acceptance +& $p_T^l$ & $p_T^{b_1}$ & $M(l_1,l_2)$& $M(l_2,b_2)$,& all\\
 Prossage & $M(l,l)$ & $<50$ GeV &$<100$ GeV, & $<70$ GeV & $M(l_1,b_2)$ & cuts \\ 
 & $\in \!\!\!\!\!/$[70,100] GeV & &$p_T^{b_2}$  & & $<75$ GeV, & in\\
 & & & $< 65$ GeV  & & $M(l_1,b_1)$ & [fb]\\
 & & & & & $<100$ GeV& \\\hline\hline
 $t \bar t + 0$-jet & 3971& 1020& 435.6& 308& 29.44& \\
 $t \bar t + 1$-jet & 668.3& 182.9& 84.23& 72.54& 8.46& 38.68\\
 $t \bar t + 2$-jet & 194.7& 70.0& 9.5& 8.57& 0.36& \\
 $bbWW$ & 50.6& 18.13& 9.31& 8.3& 0.42& \\\hline
 $t_2^1 \bar t_2^1 + 0$-jet & 3.3& 3.08& 2.65& 2.5& 1.35& \\
 $t_2^1 \bar t_2^1 + 1$-jet & 1.05& 1.00& 0.794& 0.775& 0.418& 1.898\\
 $t_2^1 \bar t_2^1 + 2$-jet & 0.65& 0.63& 0.36& 0.33& 0.13& \\\hline
 $b_2^1 \bar b_2^1 + 0$-jet & 2.81& 2.63& 1.26& 1.26& 0.34& \\
 $b_2^1 \bar b_2^1 + 1$-jet & 0.941& 0.861& 0.437& 0.437& 0.147& 0.514\\
 $b_2^1 \bar b_2^1 + 2$-jet & 0.46& 0.43& 0.114& 0.114& 0.027& \\\hline\hline
\end{tabular}
\end{center}
\caption{\it 
Effects of acceptance and different selection cuts applied on signal and background at the LHC with $\sqrt s=14$ TeV. For illustration, we have chosen $R^{-1}$ = 800 GeV for signal 
processes.}
\label{table_cuts}
\end{table}

\begin{figure}[!h]
\label{fig:lumi}
\begin{center}
\vspace*{-3ex}
\includegraphics[width=7cm,height=8.0cm,angle=-90]{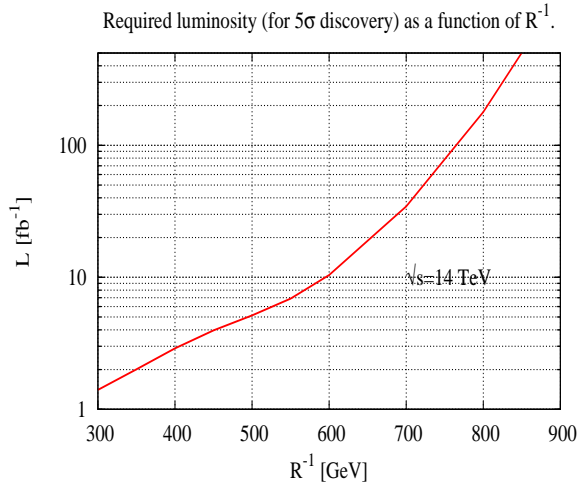}

\caption{{\it Required luminosities for $N/ \sqrt{ N + B} > 5$ 
for a LHC run with center of mass energy 14 TeV. }} 
\label{luminosity} 
\end{center}
 \end{figure}

In Table \ref{table_cuts}, we systematically list the effects of the
different cuts that we have imposed for a operating 
energy of $\sqrt{s} = 14 \tev$.
For illustration, we choose $R^{-1} = 800 \gev$.
It is clear that a large fraction of the background is eliminated 
while losing only a small part of the signal.
Although further refinements are possible and a set of cuts tuned to  
a specific value of $R$ may be imposed (thereby improving the 
signal to noise ratio even further), we desist from doing so. 
Rather, we limit ourselves to the aforementioned cuts alone and 
examine the reach of the LHC
experiments.  Quantifying the {\it statistical significance} of the 
result through the use of the ratio $S \equiv N/ \sqrt{ N + B}$ ($N$
and $B$ being the number of signal and background events for a given
luminosity), a discovery potential is nominally designated by $S > 5$.
 In Fig.\ref{luminosity}, we present the luminosity required 
to obtain a $S = 5$ result as a function of 
$R^{-1}$ at the LHC with $\sqrt s=14$ TeV. 
We conclude that, with the LHC running at a center of mass energy of 14 TeV,
a 5$\sigma$ discovery is possible upto 600 (750) GeV with an 
integrated luminosity of $10 (100) {\, \rm fb}^{-1}$. 

For the $\sqrt{s} = 7 \tev$ run, though, the expectations are far 
more modest. As Table \ref{tab:cuts_7tev} shows, the fractional 
reductions now are similar to those for the $\sqrt{s} = 14 \tev$ 
case. However, the smaller energy means that the cross sections are 
smaller. Coupled with the fact that the integrated luminosity expected 
for such a run is smaller too, this leads to a rather modest reach. 

\begin{table}[h]

\begin{center}

\begin{tabular}{||c|c||c|c||c||}
\hline \hline
\multicolumn{2}{|c||}{} & \multicolumn{2}{|c||}{Cross-section in fb} & Total after\\\cline{3-4}
\multicolumn{2}{|c||}{Process} & Acceptance Cuts + & Selection Cuts & Selection Cuts\\
\multicolumn{2}{|c||}{} & $M(l,l)\in \!\!\!\!\!/$[70,100] GeV & & [fb]\\\hline\hline
 &  $t \bar t + 0$-jet & 618.6& 6.23& \\
Background  &  $t \bar t + 1$-jet & 99.34& 1.66& 7.92\\
 &  $t \bar t + 2$-jet & 21.28& 0.03& \\\hline\hline
 &  $t_2^1 \bar t_2^1 + 0$-jet & 10.19& 5.04& \\
Signal &  $t_2^1 \bar t_2^1 + 1$-jet & 3.66& 2.01& \\
for &  $t_2^1 \bar t_2^1 + 2$-jet & 1.2& 0.32& 7.56\\\cline{2-4}
$R^{-1}=250$ GeV &  $b_2^1 \bar b_2^1 + 0$-jet & 0.36& 0.18& \\
 &  $b_2^1 \bar b_2^1 + 1$-jet & 3.2$\times 10^{-2}$& 7.2$\times 10^{-3}$& \\
 &  $b_2^1 \bar b_2^1 + 2$-jet & 1.4$\times 10^{-2}$& 1.9$\times 10^{-3}$& \\\hline\hline
 &  $t_2^1 \bar t_2^1 + 0$-jet & 4.61& 1.93& \\
Signal &  $t_2^1 \bar t_2^1 + 1$-jet & 1.74& 1.03& \\
for &  $t_2^1 \bar t_2^1 + 2$-jet & 0.58& 0.14& 3.54\\\cline{2-4}
$R^{-1}=300$ GeV &  $b_2^1 \bar b_2^1 + 0$-jet & 0.58& 0.31& \\
 &  $b_2^1 \bar b_2^1 + 1$-jet & 0.21& 0.13& \\
 &  $b_2^1 \bar b_2^1 + 2$-jet & 3.4$\times 10^{-2}$&1.9$\times 10^{-2}$ & \\\hline\hline

\end{tabular}

\end{center}
\caption{{\it
Signal and background cross-section after selection and acceptance cuts for the LHC with 7 TeV center-of-mass energy.}}
\label{tab:cuts_7tev}
\end{table}


\section{Summary and Conclusion} 

Universal extra-dimensional models, which are phenomenologically interesting
candidates for physics beyond the Standard Model, are 
characterized by towers of Kaluza Klein excitations for each of the SM
particle. In the minimal version of the UED (in which we are
interested in the present work), these KK states are characterized by
a single positive integer $n$ known as the KK number, with a 
mass-separation that is almost constant. In addition, 
the higher-dimensional theory being necessarily non-chiral, at every 
KK-level, there correspond two excitations corresponding to each SM fermion 
(one being a $SU(2)$ doublet and the other a singlet).
Conservation of KK parity stipulates both that the first-level 
states can only be pair-produced and that the 
lightest of these constitutes a stable weakly interacting massive 
particle ($\gamma^1$ in our case) and, hence, a candidate for 
the Dark Matter density of the universe. 

In this paper, we consider the pair production of the $n=1$
KK-excitations of top and bottom quarks at the LHC.  Once produced,
the heavy top and bottom quarks decay to SM $b$ quarks along with
$W^1$ and $Z^1$ (KK-excitations of gauge bosons). The decays of the
gauge boson excitations, in turn, give rise to leptons and missing
$p_T$ (from the undetected $\gamma^1$). Thus, the signal constitutes a
pair of 
tagged
$b$ jets, opposite sign di-leptons (not necessarily of the
same flavour) and missing $p_T$. This, of course, is a classic signature for
the SM top quark pair production and, hence, would be studied very well 
at the LHC.

While the $t \bar t$ (and other assorted SM) background for these
final state is much larger compared to the typical signal size, we
show that, with a careful choice of kinematic cuts, the signal can be
enhanced {\it vis-\`{a}-vis} the background, while retaining enough
signal events for a positive verdict. To be quantitative, with the
LHC operating at $\sqrt{s}$ = 14 TeV, an accumulated luminosity of
$10 (100)\, {\rm fb}^{-1}$, would allow us to discover UED in this
mode as long as $R^{-1}$ does not exceed about 600 (750) GeV. 
It should be emphasized here that our
analysis neither takes care of all experimental effects, nor is the
event selection algorithm the most optimized one. Combined with the
fact that we deliberately underestimate the signal strength by not
incorporating the QCD corrections (we include those for the SM
backgrounds though), our analysis is thus of an exploratory
nature. However, given the fact this final state would, anyway, be
well studied, the positive nature of our results encourages a detailed
and careful analysis of these modes with a realistic and full detector
simulation. For example, it should be noted that we have not 
included soft initial state radiation (ISR) effects. Typically, the production
of a pair of high mass particles would often be associated with
the emission of an associated 
high--$p_T$ jet~\cite{Plehn:2005cq, Alwall:2008qv}. The typical ISR 
for $t\bar t$ production would tend to be softer. This, at first 
sight, might seem an additional discriminant between signal and background. 
On the other hand, the emission of such hard jets would tend to change the 
$p_T$ profiles to an extent. 
However, even the inclusion of such hard 
jets maintains the essential difference between the signal 
and background profiles.
Evaluation of the entire set of effects 
needs a detailed simulation though and we reserve it for 
a future study.

Finally, it should be noted that the spectrum and the couplings in UED
bear some similarity to supersymmetric and many a different
non-supersymmetric scenarios which contain additional gauge bosons
and/or vector-like fermions.  However, in UED, the SM particles and
their KK partners share the same spin. This is in contrast to the
supersymmetric (which may though be argued of as an extra dimension in
fermion coordinates) extension of the SM, wherein the SM particles and
their superpartners carry different spins. We would like to emphasize
that, from an experimental point of view, UED is closer to
supersymmetry. (as supersymmetry with conserved $R$-parity does
contain a stable weak massive particle).  To be more specific, pair
production of top-squarks in MSSM and their consequent decay would
produce the same signal we have discussed above.  Differentiating
these two new physics alternatives will have to rely on the
dissimilarity in final state phase space distributions on account of
the difference in the spins of particles being produced. It might be
argued that for a $t^1_2$ and a top-squark of identical masses, the
UED scenario would, typically, be blessed with a significantly higher event
rate. While this is indeed true, the fraction of events that survive
the cut is a very strong function of the mass-splitting, and in the
case of supersymmetric theories, depends crucially on the details of
the model.  UED models, on the other hand, are relatively free of such
model-dependence. 
Although the quantum corrections
to the spectrum do depend upon the ultra-violet completion of 
the theory, the fractional splitting of the masses 
within a generation typically remains small. In particular, the nature of 
the $t_2^1$ and $b_2^1$ cascades, generically, remain similar 
to those we considered here, notwithstanding 
some model-dependence.
 A more quantitative study of how to differentiate
UED from supersymmetry, following our line of analysis in the context
of LHC, is beyond the scope of the present work.

\vskip 5pt
{\bf Acknowledgements}:~  We thank Peter Skands for some 
very insightful comments. 
DC acknowledges support from the Department of
Science and Technology, India under project number SR/S2/RFHEP-05/2006. 
The research of  AD is partially
supported by project no.~2007/37/9/BRNS (DAE), India and 
by CSIR project no. 03(1085)/07/EMR(II). KG acknowledges support from 
CSIR, India.

\vskip 5pt

\end{document}